\pdfoutput=1
\documentclass[cits]{JINST}

\usepackage{booktabs}
\usepackage{subfigure}
\usepackage{textcomp}

\title{Thin silicon strip detectors for beam monitoring in Micro-beam Radiation Therapy}

\author{M.~Povoli$^a$\thanks{Corresponding author.}, 
E.~Alagoz$^b$, 
A.~Bravin$^c$, 
I.~Cornelius$^d$, 
E.~Br\"auer-Krisch$^c$, 
P.~Fournier$^d$, 
T.~E.~Hansen$^e$, 
A.~Kok$^e$, 
M.~Lerch$^d$, 
E.~Monakhov$^a$, 
J.~Morse$^c$, 
M.~Petasecca$^d$, 
H.~Requardt$^c$, 
A.~B.~Rosenfeld$^d$, 
D.~R\"ohrich$^b$, 
H.~Sandaker$^f$,
M.~Salom\'e$^c$, 
B.~Stugu$^b$.\\
\llap{$^a$}Centre for Material Science and Nanotechnology, University of Oslo,\\
Postboks 1048 Blindern 0316 Oslo, Norway\\
\llap{$^b$}Department of Physics and Technology, University of Bergen,\\ 
Postboks 7803, N-5020 Bergen, Norway\\
\llap{$^c$}European Synchrotron Radiation Facility, 71 avenue des Martyrs, 38000 Grenoble, France\\
\llap{$^d$}Centre for Medical Radiation Physics, University of Wollongong, NSW 2522, Australia\\
\llap{$^e$}SINTEF MiNaLab, Department of Microsystems and Nanotechnology,\\ 
Gaustadall\'een 23 C, Oslo, Norway\\
\llap{$^f$}Depatment of Physics, University of Oslo, Postboks 1048 Blindern 0316 Oslo, Norway\\

E-mail: \email{marco.povoli@smn.uio.no}}

\abstract{
Microbeam Radiation Therapy (MRT) is an emerging cancer treatment that is currently being developed at the 
European Synchrotron Radiation Facility (ESRF) in Grenoble, France.  This technique uses a highly collimated 
and fractionated X-ray beam array with extremely high dose rate and very small divergence, to benefit from the 
dose-volume effect, thus sparing healthy tissue. In case of any beam anomalies and system malfunctions, 
special safety measures must be installed, such as an emergency safety shutter that requires continuous 
monitoring of the beam intensity profile.  
Within the 3DMiMic project, a novel silicon strip detector that can tackle the special features of MRT, such 
as the extremely high spatial resolution and dose rate, has been developed to be part of 
the safety shutter system.  The first prototypes have been successfully fabricated, and experiments aimed to 
demonstrate their suitability for this unique application have been performed. Design, fabrication and the 
experimental results as well as any identified inadequacies for future optimisation are reported and discussed in 
this paper.
}


\graphicspath{ {./figures/} }

\begin{document}


\section{Introduction}
Micro-beam Radiation Therapy (MRT) is an emerging cancer treatment first proposed 
at the Brookhaven National Laboratory (BNL), USA \cite{Slatkin95}, and later developed at the 
European Synchrotron Radiation Facility (ESRF), Grenoble, France \cite{Krisch09}. 
The technique relies on the irradiation of tumors using an array of parallel X-ray micro-beams 
generated by a Multi-Slit Collimator (MSC). Each beam is a few tens of microns wide (selectable between
25 and 100~$\mu$m) with a centre-to-centre (c-t-c) distance of between 100 and 400~$\mu$m. 
Third generation synchrotron facilities have a negligible beam divergence and very large dose rates.
The X-ray energies involved in the MRT span from 27 to 600~keV and the delivery of irradiation dose rates 
can be as high as 20~kGy/s.
These properties are essential for MRT in order to minimise the treatment time and to avoid the lateral
spreading of the micro-beams caused by cardiosynchronous movement of tissues.
The use of the MRT beam geometry to deliver large radiation doses has shown many promising results
in preclinical trials on different animal models, including mice, rats, piglets, and rabbits
\cite{Slatkin92,Kim93,Laissue98,Schultke08,Serduc09}.

Due to the unique shape of the fractionated beam, the irradiation in tissues consists of peaks (high dose) 
and valleys (low dose). One crucial quantity to be monitored before and during the MRT treatment
is the so-called Peak to Valley Dose Ratio (PVDR). The highest possible PVDR is of great radiobiological 
importance. The high dose or "peak region" is essential in destroying the tumour cells while the low dose or "valley 
region" must be sufficiently low such that healthy tissue can repair for promising recoveries following the treatment.
A real-time measurement of the PVDR (with an accuracy of 3\% or better) in a short time frame 
(microsecond timescale), is therefore of great importance for effective treatment planning and 
dosimetry quality assurance. 

Different types of dosimeters have been tested for MRT including:
edge-on MOSFET devices \cite{Rosenfeld99,Rosenfeld01,Krisch03,Siegbahn09}, 
flash memory MOSFETs \cite{Cellere04}, MRI gel dosimeters \cite{Bayreder06,DeDeene02}, fluorescent 
nuclear track detectors \cite{Sykora10}, and high resolution TLD dosimeters \cite{Ptaszkiewicza08}.
Promising results have been demonstrated on these devices but most of them lack the ability for a fast real time 
online measurement or are not sufficiently radiation hard. Both are crucial requirements for 
treatment planning and quality assurance in MRT.  

Silicon dosimeters can easily be coupled to an on-line readout electronic system. 
Together with their high spatial resolution ability down to a few micrometers, this makes silicon devices a prime 
potential candidate as a beam monitor for MRT.  
The first investigated silicon dosimeter for MRT consisting of a single strip (or single sensing element) was 
proposed and designed by the Center for Medical Radiation Physics (CMRP) at the University of 
Wollongong in Australia \cite{Lerch11}. Encouraging results in terms of dose linearity, spatial resolution 
and radiation tolerance were observed.
In order to monitor the entire array of microbeams in MRT simultaneously, multiple sensing 
elements must be implemented. This paper concerns such multi-strip silicon detectors that were fabricated 
at SINTEF, MiNaLab (Oslo, Norway) within the framework of the 3D MiMic project \cite{3DMiMic}.
The fabrication process from design to electrical characterisation and experimental 
testing of the detectors using relevant X-ray beam-lines are reported here. 
The experiments performed aimed to demonstrate the sensor suitability for MRT and to identify 
the design and fabrication modifications for future optimisation.

\section{Thin silicon beam monitors for the MRT}\label{sec:beam_monitor}
Previous experimental testing on the single strip sensors at the biomedical beam-line ID17 at the 
ESRF \cite{Lerch11} provided a good insight into the necessary elements and design parameters to be 
modified and improved, when one implements the design for a multi-strip silicon detector to be part of a 
reliable monitoring system for the MRT. 
Due to the unique properties of MRT beam, the requirements for both the single 
strip and the multi-strip sensors differ greatly from those of conventional planar silicon devices. 
The device requirements for the desired sensors in MRT include:
\begin{itemize}
\item{near zero beam perturbation during treatment;}
\item{no signal saturation in the readout electronics;} 
\item{near zero cross-talk between neighbouring channels;}
\item{large total sensitive area with high spatial resolution;}
\item{radiation hardness when exposed to the MRT X-ray beam array.}
\end{itemize}

\begin{figure}[!t]%
\centering 
\subfigure[]{\label{fig1a} 
	\includegraphics[width=.225\textwidth]{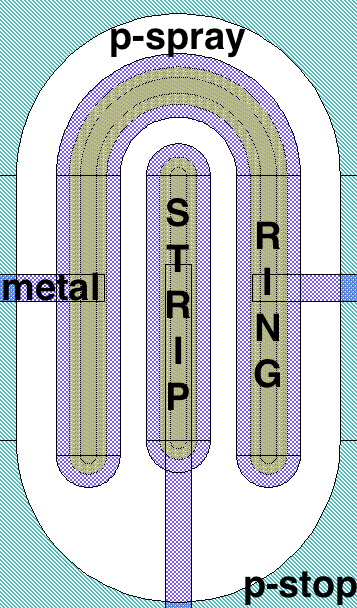}
}%
\subfigure[]{\label{fig1b} 
	\includegraphics[width=.225\textwidth]{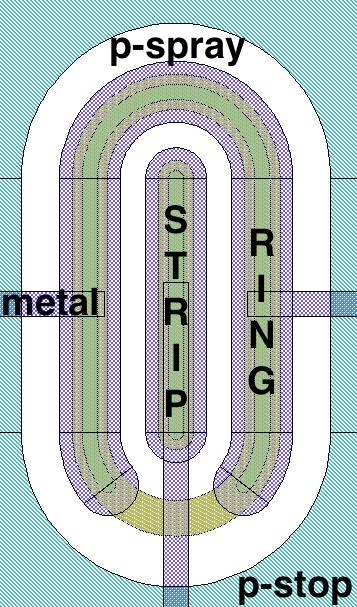}
}%
\subfigure[]{\label{fig1c} 
	\includegraphics[width=.365\textwidth]{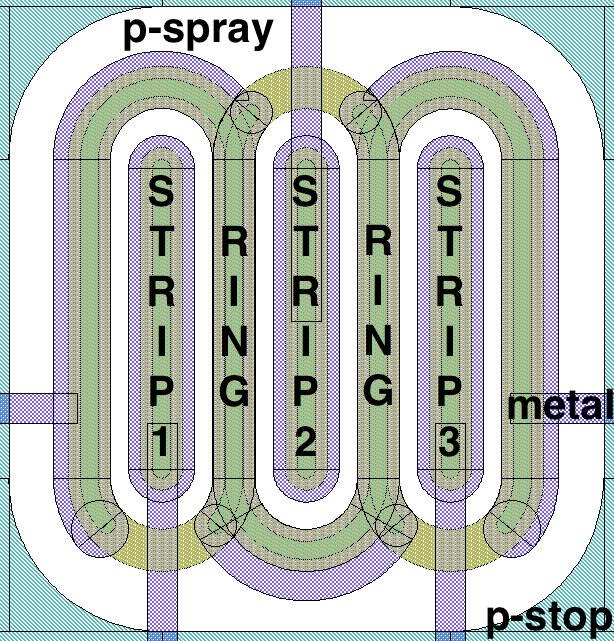}
}%
\caption{Standard layout for a 50~$\mu$m long sensitive element: (a) strip with an open steering-ring, 
(b) strip with a closed steering-ring and (c) three strips with reduced pitch and closed steering-ring for the 
variable pitch design.} 
\label{fig1}
\end{figure}

\subsection{Device layout}
The sensor design was carried out based on the above requirements, 
and was later validated with numerical simulations (see subsection \ref{sub:numerical_sim}).
The layout solutions for each of the identified requirements can be summarised as follows.
\begin{enumerate}
\item{Minimal beam perturbation is achieved by the fabrication of thin sensors,
approximately 10~$\mu$m thick. This reduces the total active volume available 
for ionisation, and thus results in smaller generated signals that minimise the saturation of the overall 
readout system.
Details on the thinning of the silicon detectors are reported in section \ref{sec:fabrication}.} 
\item{In order to ensure the required spatial resolution over a large active area and to reduce the 
measurement time for the full radiation field, it is necessary to produce a sensor with multiple sensitive elements. 
A strip sensor is deemed to be the optimal choice for this application. The strip lengths must be kept short
to limit the sensitive volume of each element, in the range of 10 to 250~$\mu$m.}
\item{To limit the amount of charge reaching the sensitive elements, a "steering-ring" has been
implemented around each strip. The steering ring sinks the excess signal to ground (see Figure~\ref{fig1}). 
The use of the steering-ring is also expected to eliminate cross-talk and charge sharing between neighbouring
channels, effects that were found to be detrimental to the correct measurement of the PVDR and FWHM
of the single micro-beam \cite{Kalliopuska11}. }
\item{Due to the extreme radiation doses expected in the MRT, it is important to find solutions to mitigate the
radiation damage. Because of the photon energies involved in this treatment, the damages will mostly be 
localised in the surface layers (increase of the oxide charge and trapping centres at the Si/SiO$_2$ interface). 
To limit the changes in the device behaviour, the surface layers must be designed using both p-spray and p-stop 
implantations \cite{Piemonte06}. This is a well-known technique used to compensate the increase in oxide
charge and to retain the inter-channel isolation in position sensitive silicon detectors. 
A small amount of bulk damage is also expected, but the performance degradation due to this damage is
insignificant or negligible.}
\end{enumerate}

In order to gain an in-depth understanding and to identify the most suitable sensor, many different 
geometrical implementations were studied. 
The standard strip pitch was set to 100~$\mu$m and the strip length varied between 10 and 250
$\mu$m (the optimal value will be chosen after testing the sensors under MRT conditions).
Two different implementations of the steering-ring were also included: in one case the ring 
was designed as an open implantation (Figure~\ref{fig1a}), while in the second case the ring implant 
was closed to ensure full strip isolation (Figure~\ref{fig1b}). 
An additional variation, intended to locally increase the sampling on each micro-beam 
was also included. In this design, a group of three strips (strip triplet) was placed alternately between 
single strips at a pitch of 200~$\mu$m while the pitch within the triplet was 24~$\mu$m (Figure~\ref{fig1c}). 
Each strip in the triplet is fully enclosed individually by the steering ring.
All sensors have 128, 250 or 400 channels, depending on the width of the 
micro-beam array to be monitored. Devices were fabricated on p-type substrates with three different wafer 
resistivities: 5, 100 and  10$^4$~$\Omega$cm.
The use of different substrates was to investigate if the corresponding recombination rate can further reduce 
the risk of readout saturation under full intensity MRT beam conditions. 

Due to the large variations in design geometry and fabrication parameters, this paper will only discuss the 
results from the most relevant and appropriate sensor design.

\subsection{Numerical simulations}\label{sub:numerical_sim}
\begin{figure}[!t]%
\centering 
\subfigure[]{\label{fig2a} 
	\includegraphics[width=.45\textwidth]{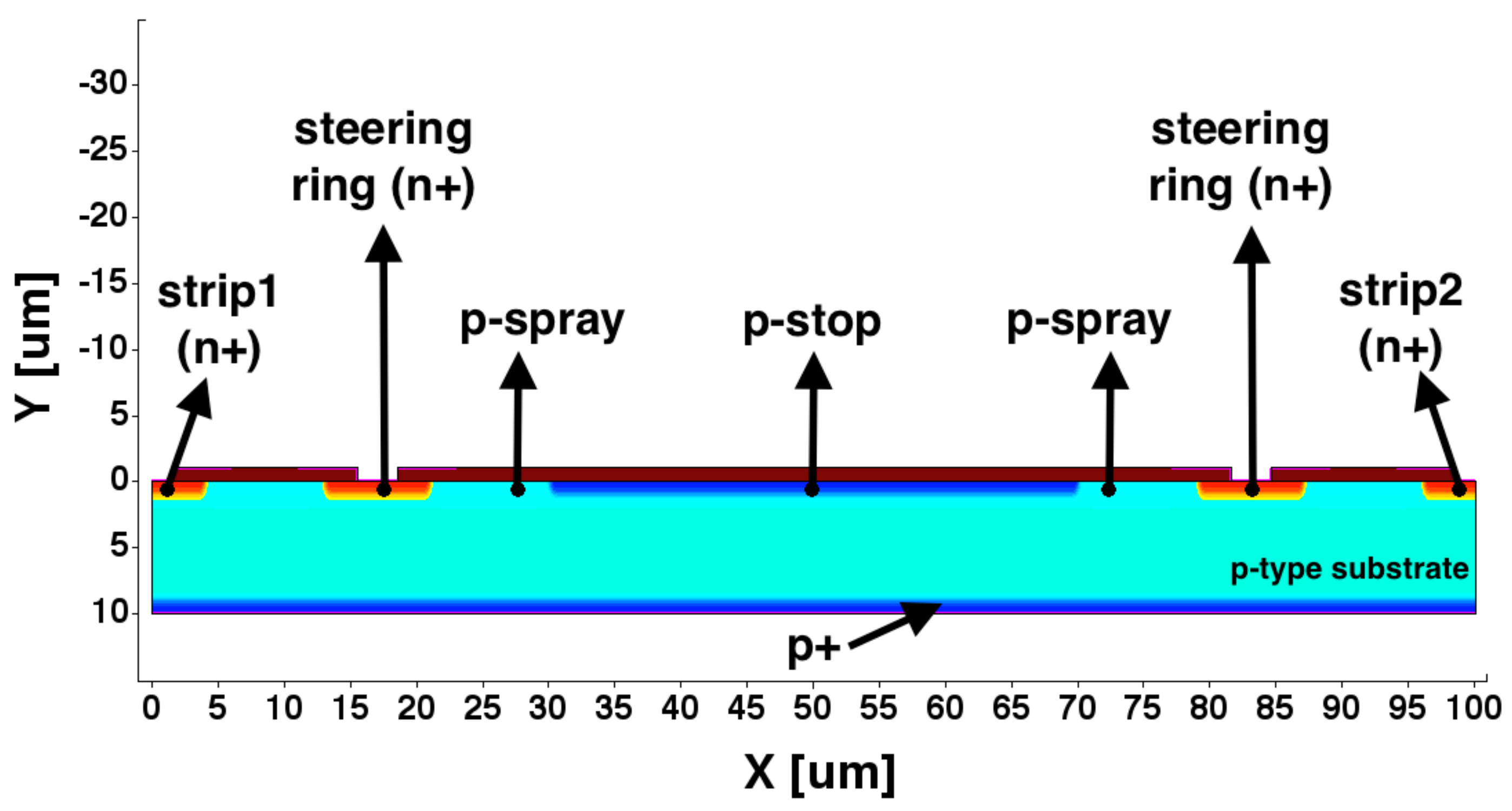}
}%
\subfigure[]{\label{fig2b} 
	\includegraphics[width=.45\textwidth]{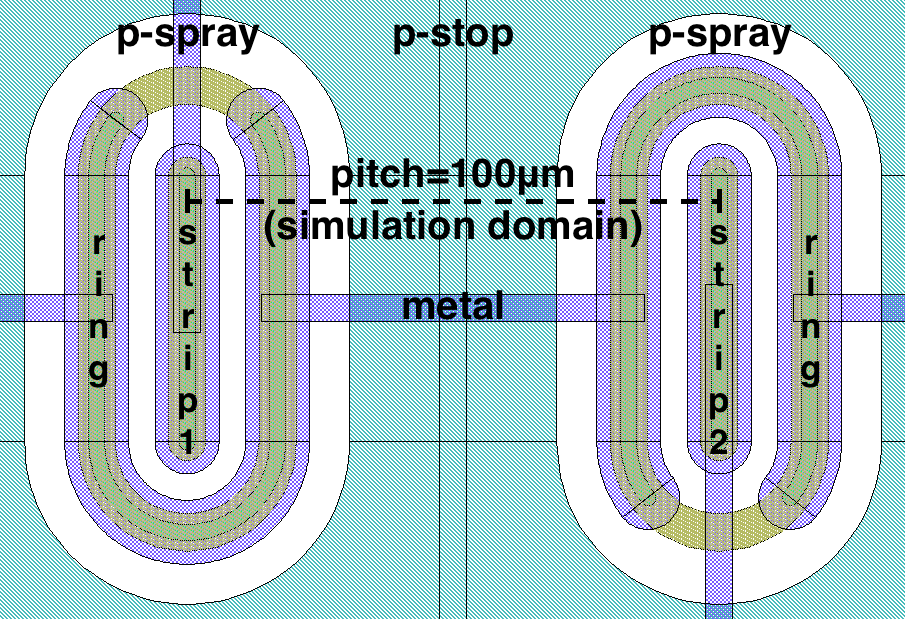}
}%
\caption{Simulated structure (a) and corresponding simulation domain on the device layout (b).} 
\label{fig2}
\end{figure}

The various layout options were validated during the design phase by means of numerical
simulations using the Synopsys TCAD tools \cite{TCAD-Synopsys}. 
Exploiting the repetitive patterned structure of the devices, it was possible to produce an 
optimised numerical grid with a reduced number of nodes, allowing faster computation. 
An example of a simulated structure is shown in Figure~\ref{fig2a}. This structure is a 2D cross-section
of the simulation domain shown in Figure~\ref{fig2b}. Two different bulk doping concentrations were investigated, 
1$\times$10$^{15}$ and 1$\times$10$^{14}$~at.~B/cm$^3$. The p-spray doping profile was extracted from
process simulations, while all the other profiles were created using 
Gaussian distributions with peak values of 5$\times$10$^{19}$~cm$^{-3}$ and junction depths of 
$\sim$2~$\mu$m.
The performed simulations focused on obtaining electrical and charge collection characteristics
of the investigated structures.
In all cases the reverse voltage was applied on the backside contact (p$^+$) and the currents (or capacitances)
were readout from the front side contacts (n$^+$). All relevant quantities were scaled
to represent a reference strip length of 50~$\mu$m.
\begin{figure}[!b]%
\centering 
\subfigure[]{\label{fig3a} 
	\includegraphics[width=.4\textwidth]{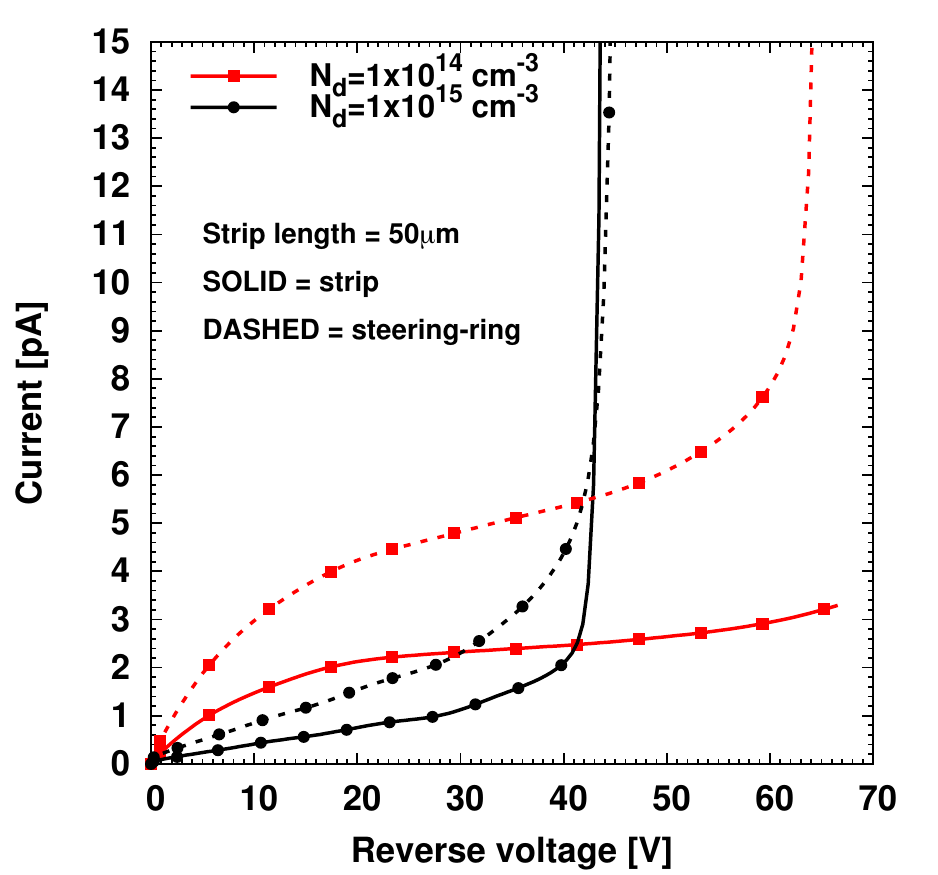}
}%
\subfigure[]{\label{fig3b} 
	\includegraphics[width=.4\textwidth]{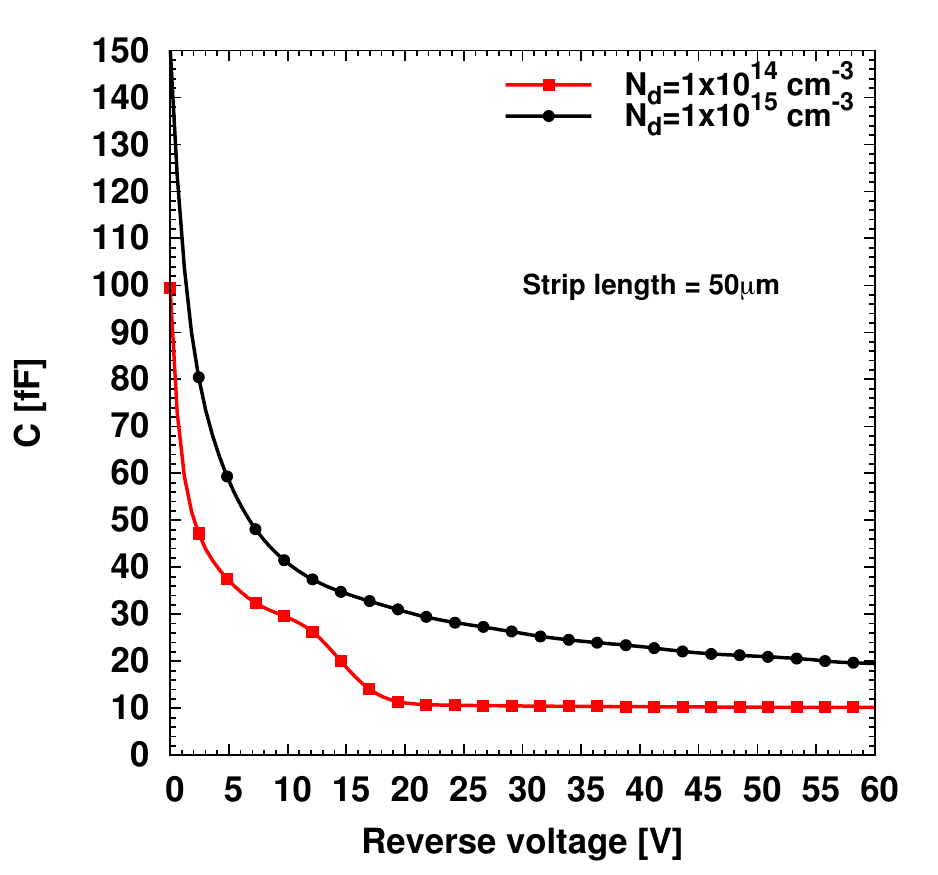}
}%
\caption{Results of the electrical simulations for a strip length of 50~$\mu$m and for two different 
bulk doping concentrations. Reverse currents (a) and capacitances (b) are reported.} 
\label{fig3}
\end{figure}

\subsubsection{Current-Voltage (I-V) simulations}
\begin{figure}[!t]%
\centering 
\subfigure[]{\label{fig4a} 
	\includegraphics[width=.6\textwidth]{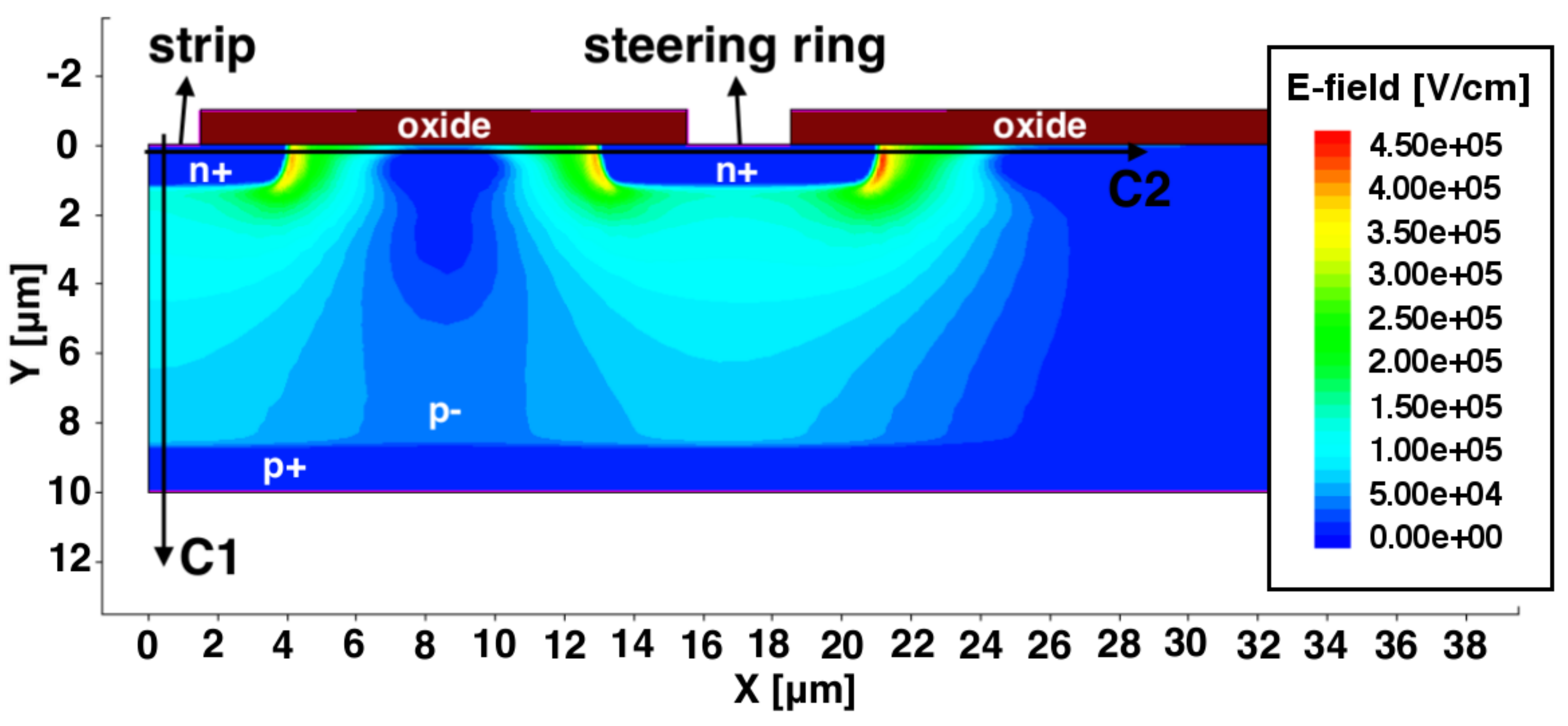}
}\\%
\subfigure[]{\label{fig4b} 
	\includegraphics[width=.4\textwidth]{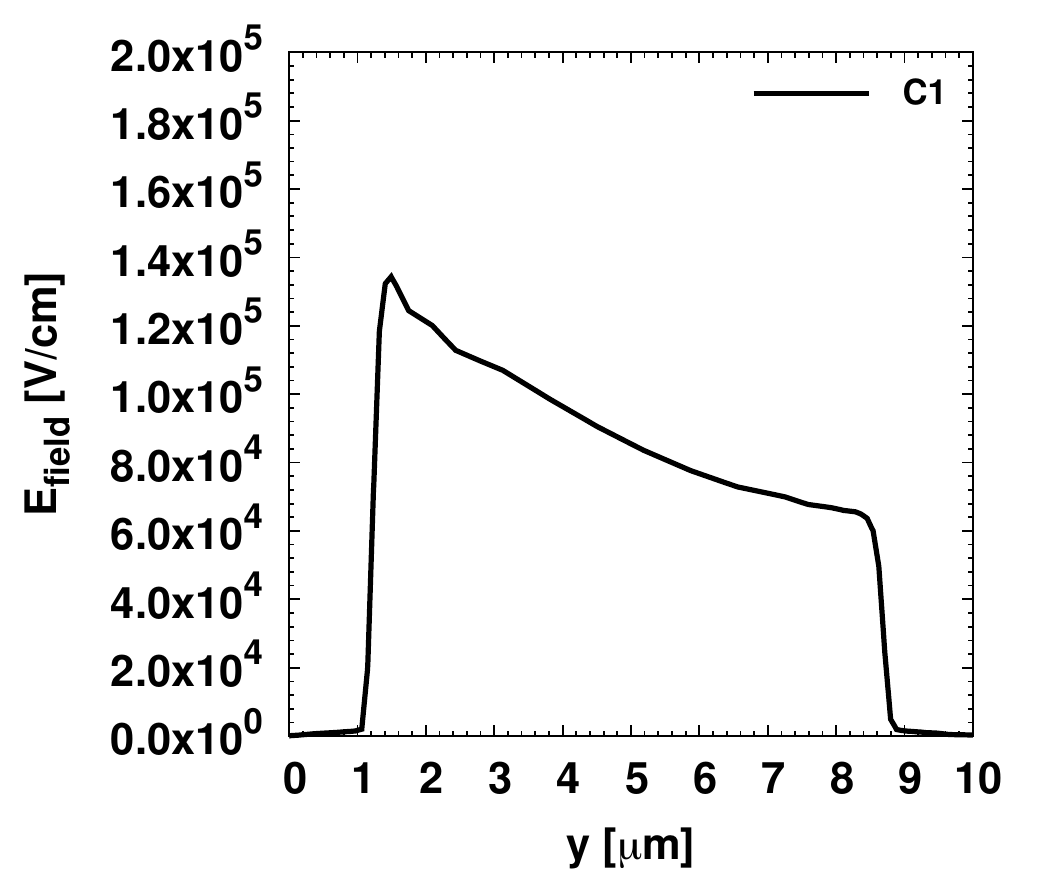}
}%
\subfigure[]{\label{fig4c} 
	\includegraphics[width=.4\textwidth]{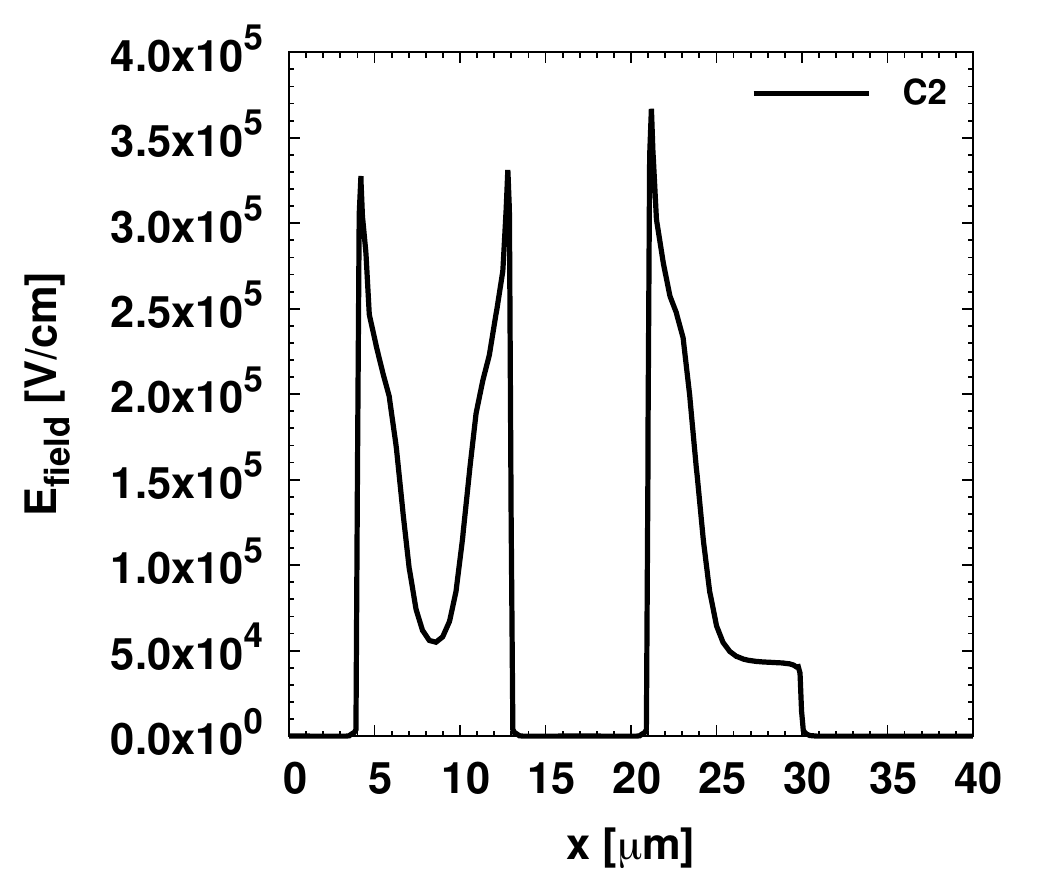}
}%
\caption{Simulated electric field distribution at breakdown for a device having a bulk doping concentration
of 1$\times$10$^{14}$~at.~B/cm$^3$. 2D map (a) and 1D field profiles extracted from (b) underneath the strip 
and (c) from the upper surface are shown.} 
\label{fig4}
\end{figure}

The simulated I-V curves are shown in Figure~\ref{fig3a} and clearly show that the expected reverse bias
current on the single strip is small, in the range of 1-2~pA. 
The steering ring exhibits a larger current because of its larger 
active volume. Contrary to expectation, the current is lower for the highest bulk doping concentration, 
as the highly doped sensor never reaches full depletion before breakdown. 
If a high enough bias voltage could be applied, the current of the device with higher bulk 
doping concentration would eventually become larger than that of the lightly doped sensor.
The breakdown voltage is rather low in both cases, due to the high doping concentration of the p-spray 
implant on the front side. This can be understood from the simulated electric field distribution inside 
the structure shown in Figure~\ref{fig4a}. The field under the strip (cutline C1, Figure~\ref{fig4b}) has the classic 
shape of a fully depleted diode, while large peaks are found close to the front surface
(cutline C2, Figure~\ref{fig4c}). These peaks are located at the intersection point between the p-spray and
the n$^+$ implant.
Finally, a lower bulk doping concentration seems to suggest an increase in breakdown voltage of roughly 20~V. 

\subsubsection{Capacitance-Voltage (C-V) simulations}
Capacitance simulations are shown in Figure~\ref{fig3b}. Only the dominant strip to back-side capacitance is
reported here, since the strip to steering-ring parasitic capacitance is negligible. To facilitate the simulation and
avoid convergence problems, avalanche models were disabled for the C-V simulations. 
The simulated curves of two different resistivities have similar shapes at low voltages but then begin to differ 
at around 10 V. 
This is caused by the inability to reach full depletion in the higher bulk concentration device, causing 
its capacitance to not saturate completely (note that despite the different bulk doping concentrations, 
the two devices should have similar full depletion capacitance values). 

Additional information can be obtained by focusing on the curve for the device with the lower bulk doping 
concentration. The change in slope between 5 and 15~V occurs when the depletion regions of the strip and 
the ring overlap each other. The simulated capacitance values are low (10-20~fF) as expected due to the 
small active volumes of the sensitive elements.

\begin{figure}[!t]
   \centering
   \includegraphics[width=.5\textwidth]{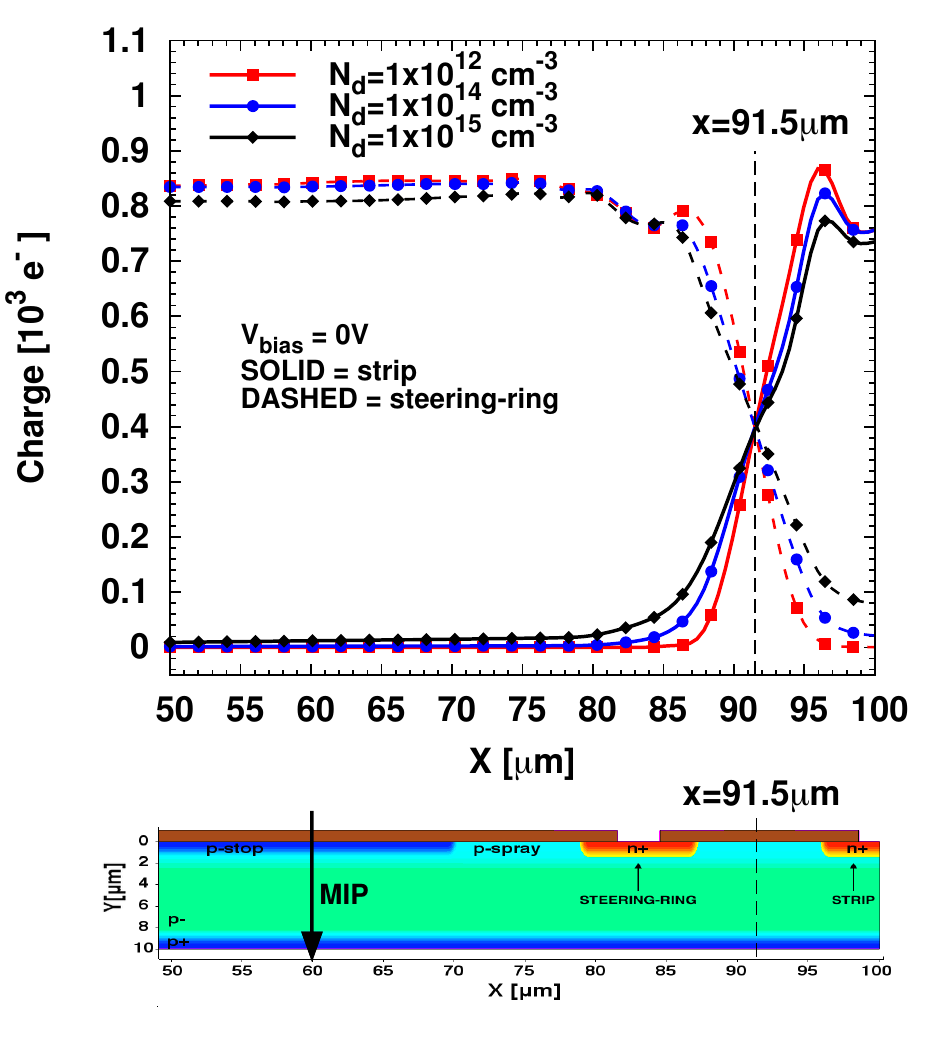}
   \caption{Simulated charge collection efficiency from a MIP impinging the structure in different positions.}
   \label{fig5}
\end{figure}
\subsubsection{Charge collection simulations}
The investigation of charge collection behaviour was performed by means of numerical
simulations in transient mode. Since the simulator does not provide an X-ray generation model, the "Heavy-Ion"
model was used. 
For this particular experiment, the aim was to extract the intrinsic charge collection dynamics 
of the sensor, without the influence of any other additional physical effects related to X-ray beam absorption and
scattering processes. A Minimum Ionising Particle (MIP) was used, generating 80~electrons/$\mu$m along
a pre-defined particle track. 
The particle was scanned across the structure in steps of 2~$\mu$m starting from $x$=50~$\mu$m 
(in the middle between two channels) and reaching $x$=100~$\mu$m (middle of the strip).
Both the steering-ring and the strip were readout separately. The current of each channel was integrated over
time and the collected charge was then plotted as a function of the hit positions for all investigated bulk 
doping concentrations. Note that the intended use of these devices is in passive mode (i.e. without external 
bias voltage), to further avoid readout saturation by the increased recombination of the generated 
charge. In addition, passive mode allows partial mitigation of the effects of surface radiation 
damage \cite{Moll_thesis}, thus extending the sensor lifetime. 

Simulation results are reported in Figure~\ref{fig5} for a bias voltage of 0V.
It can be observed that the ring collects all the charge generated for $x$<85~$\mu$m, confirming its 
correct operation. There is then a region (85<x<95~$\mu$m) where the charge is shared between 
the ring and the strip, with the crossing point located at $x$=91.5~$\mu$m. 
Reductions in charge collection are clearly visible under
the ring ($x$=82.5~$\mu$m) and under the strip ($x$=100~$\mu$m). 
This is a result of the presence of the two n+ implantations that cause recombination of the generated charge. 
This situation is critical only if the generated charge is small.
Due to the extremely high signal available for detection in the beam array of MRT, such small signal loss does 
not pose a problem.
Another important aspect highlighted by the simulations is the different amount of collected charge for 
different bulk doping concentrations, with the lowest concentration achieving highest charge collection 
as expected. 
Despite the agreement with expectations, the difference in charge collection between the different bulk
doping concentrations, does not seem to indicate any considerable advantage of one with respect to
the others. Functional testing with different radiation sources will be essential in deciding the 
best geometrical implementation of the device and the ideal bulk doping concentration for the 
MRT beam monitors (see section \ref{sec:functional-charact}).

\begin{figure}[!b]%
\centering 
\subfigure[]{\label{fig6a} 
	\includegraphics[width=.49\textwidth]{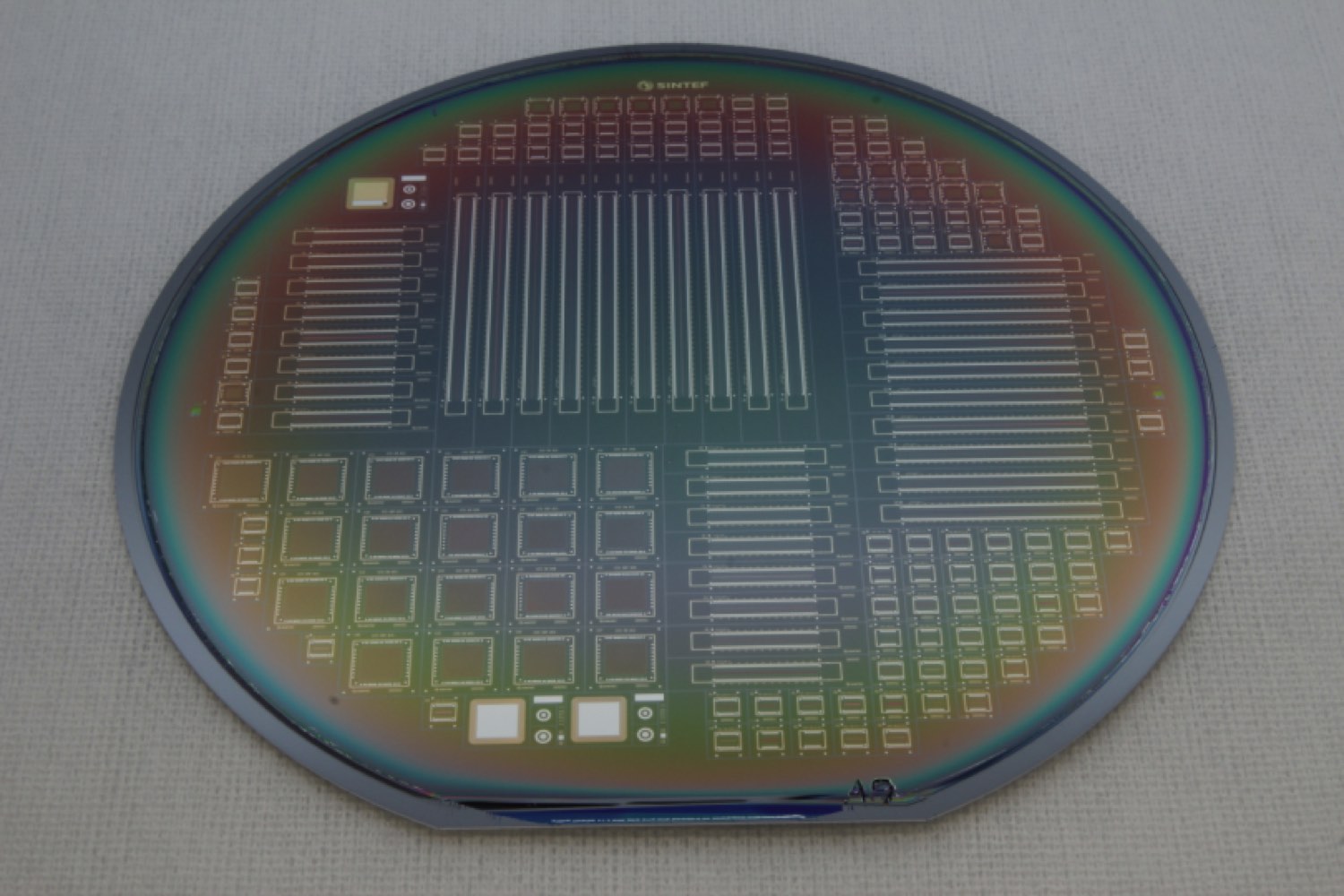}
}%
\subfigure[]{\label{fig6b} 
	\includegraphics[width=.49\textwidth]{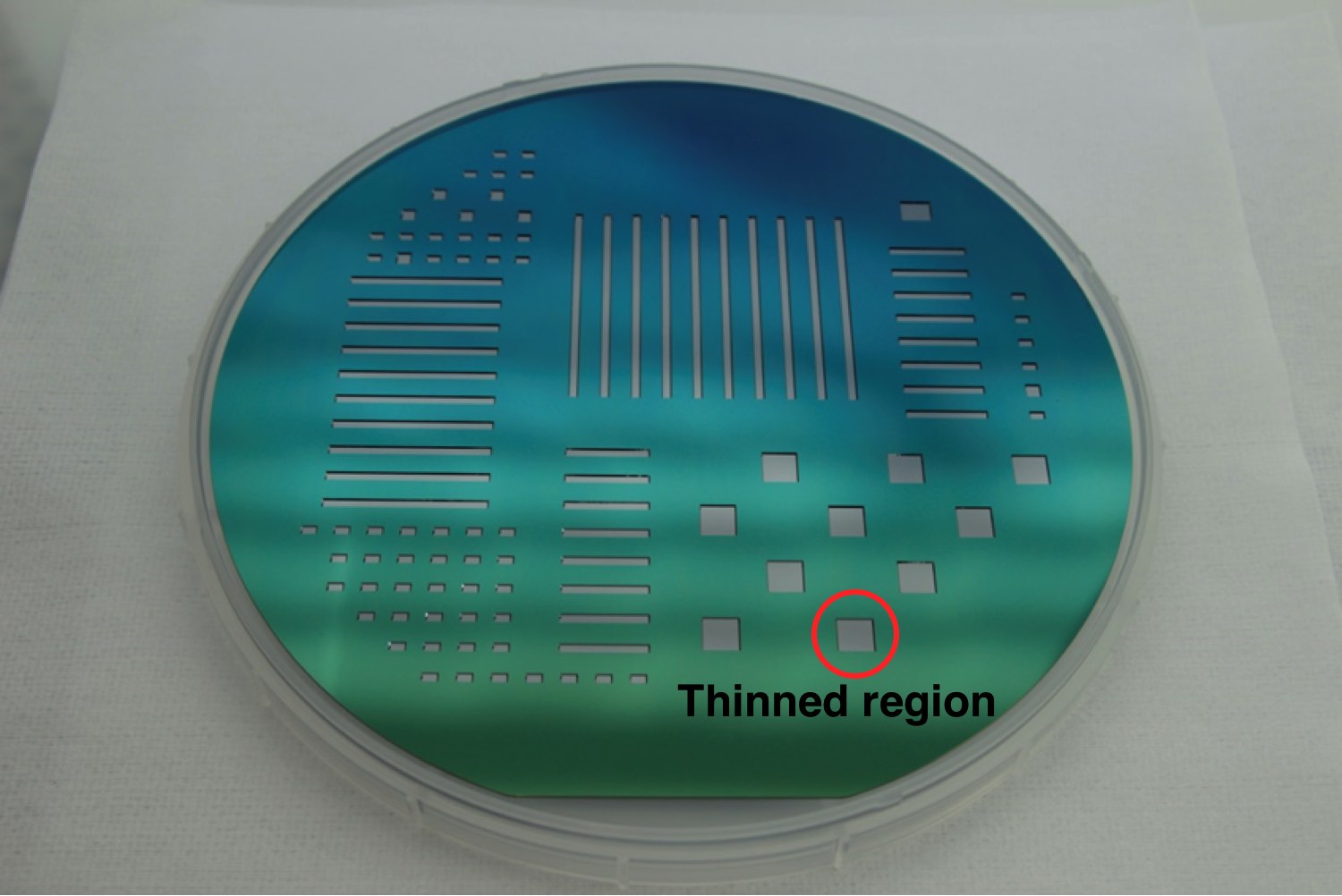}
}%
\caption{Pictures of a completed wafer. Front side (a) and back side (b).} 
\label{fig6}
\end{figure}
\section{Fabrication}\label{sec:fabrication}
The fabrication process was based on a conventional planar sensor process, 
with the addition of two p$^+$ implantations to obtain the p-spray and p-stop layers needed to counteract 
the surface radiation damage.
The only complication was the required thinning step to achieve the desired 10~$\mu$m thickness. 
The thickness reduction was performed locally on the active area of the sensors only, leaving behind 
a physical frame where the original substrate thickness was retained to sustain mechanical stability.
The thinning procedure was carried out by an anisotropic silicon etching using a wet chemical etchant, TMAH. 
The etching process is well-controlled and a high uniformity of the overall thickness can be achieved 
\cite{TMAH}. 
A layer of silicon dioxide was used as the etching mask. The extra thermal oxidation and how this affects the 
critical p-spray and p-stop implantation was investigated prior to fabrication  using Silvaco TCAD tools. 
Particularly critical was the final heat treatment required to grow 
the thick oxide layer used as protection during the TMAH etching. 
This oxidation requires the wafers to stay at high temperature for a long time, causing 
diffusion of the implanted dopants. 
This aspect would normally not be critical for sensors with a standard thickness but, in 10~$\mu$m
thick devices, a doping profile diffusing several micrometers can affect the sensor behaviour. 
Pictures of a completed wafer are shown in Figure \ref{fig6}. On the front side (Figure~\ref{fig6a}) the patterning 
of all the different sensor layouts is visible, with long rectangular devices being the MRT strip detectors. 
Around the wafer edge, several test structures are placed together with sensors for other applications. 
On the back side (Figure~\ref{fig6b}) the only visible structures
are the thinned regions behind the devices. 
The final membrane thickness was measured using a profilometer probe. The probe was first calibrated 
using a fixed silicon sample on the measurement set up.  The membrane to be measured was then placed 
directly on top of the silicon sample with the use of a wafer holder. The difference in height of the probe 
when in contact with the silicon sample and when in contact with the membrane provided an accurate thickness 
measurement.
The achieved thicknesses for a subset of the processed wafers are reported in Table \ref{tab1}.
For the low resistivity wafers (A1 to A15) the final thicknesses were
between 9.9 and 12.4~$\mu$m, very close to the design value. For the high resistivity wafers, only starting 
substrates of 500~$\mu$m thick were available at the time. The significantly thicker substrate compared to 
the low resistivity materials implies a much longer etching time which resulted in over-etching and small holes 
and damages on the wafer were subsequently created. This damage was however only local, with the majority 
of the devices still functional on the affected wafers.

\begin{table}[t]
\caption{Summary of the final etched thicknesses for a subset of the processed wafers.}
\vspace{0.25cm}
\begin{center}
\begin{tabular}{lcccccccc}
\toprule
\textbf{Wafer} & \textbf{A1} & \textbf{A2} & \textbf{A5} & \textbf{A9} & \textbf{A11} & \textbf{A15} & 
\textbf{A17} & \textbf{A19}\\
\midrule
\textbf{Bulk doping [at.~B/cm$^{-3}$]} & 10$^{15}$ & 10$^{15}$ & 10$^{15}$ & 10$^{14}$ & 
10$^{14}$ & 10$^{14}$ & 10$^{12}$ & 10$^{12}$ \\
\midrule
\textbf{Measurement 1 [$\mathbf\mu$m]} & 9.1 & 9.6 & 9.8 & 10.4 & 10.5 & 10.8 & 49.2 & 49.7\\
\textbf{Measurement 2 [$\mathbf\mu$m]} & 11.1 & 10.2 & 11.1 & 14.3 & 13.5 & 12.8 & 52.0 & 50.7\\
\midrule
\textbf{Average [$\mathbf\mu$m]} & 10.1 & 9.9 & 10.5 & 12.4 & 12.0 & 11.8 & 50.6 & 50.2\\
\bottomrule
\end{tabular}
\end{center}
\label{tab1}
\end{table}%

\section{Electrical characterisation}\label{sec:electrical-charact}
The electrical characterisation was performed on the completed wafers using a manual probe station.
The selection of good sensors was determined by the current measurement of the steering-rings.
Since all the rings are connected through a metal line and read out through a single pad,
the measurement of all the steering rings provides a good evaluation of the current in the entire sensor. 
Furthermore, all strips are also biased via the punch-through mechanism \cite{PT}. The current 
in both the strips and the ring can be measured simultaneously, making this method very reliable in detecting 
bad sensors. 
Similarly, the capacitance measurements were performed using the steering-ring, 
mainly due to the fact that the single strip capacitance is far too low to be measured reliably.
The measured values, of both current and capacitance, are scaled to represent the 
active volume of a single 50~$\mu$m long strip, in order to easily compare sensors with different strip lengths.
A more in-depth understanding of the sensor operation was obtained from the comparison of 
measurements and additional 3D numerical simulations, as discussed in the following subsections. 
The measured sensor thicknesses were used in the simulations together with the simulated
p-spray implantation, for the best possible representation of the measured results. 

\begin{figure}[!t]%
\centering 
\subfigure[]{\label{fig7a} 
	\includegraphics[width=.45\textwidth]{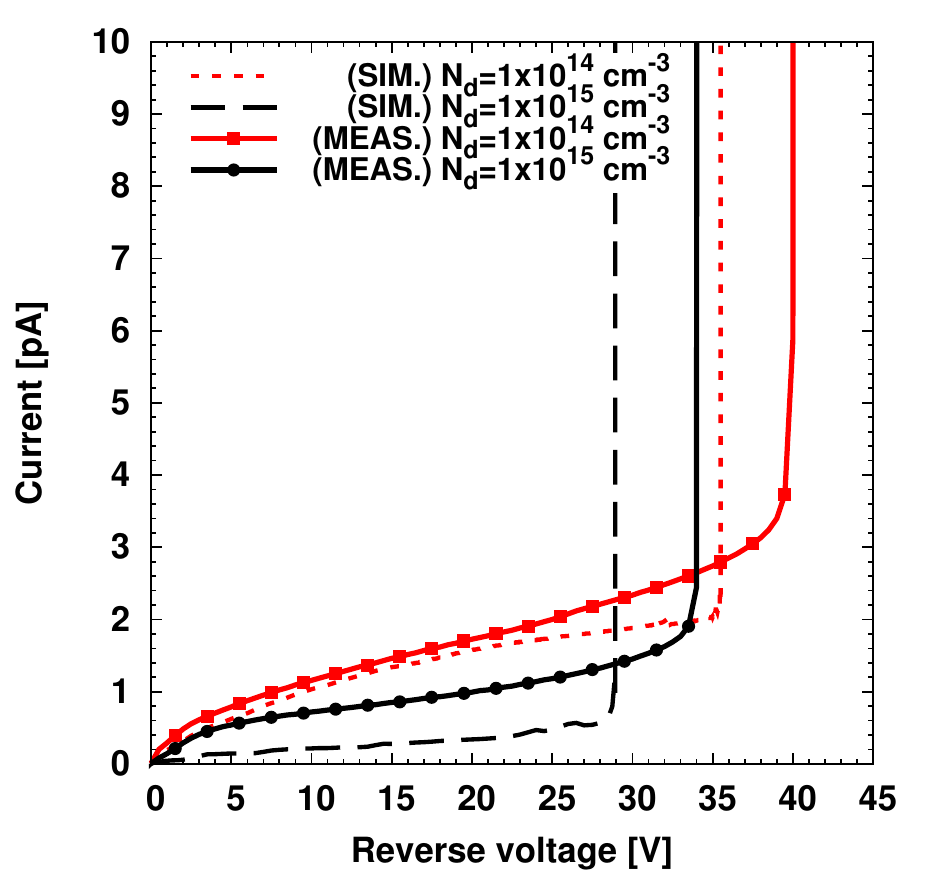}
}%
\subfigure[]{\label{fig7b} 
	\includegraphics[width=.45\textwidth]{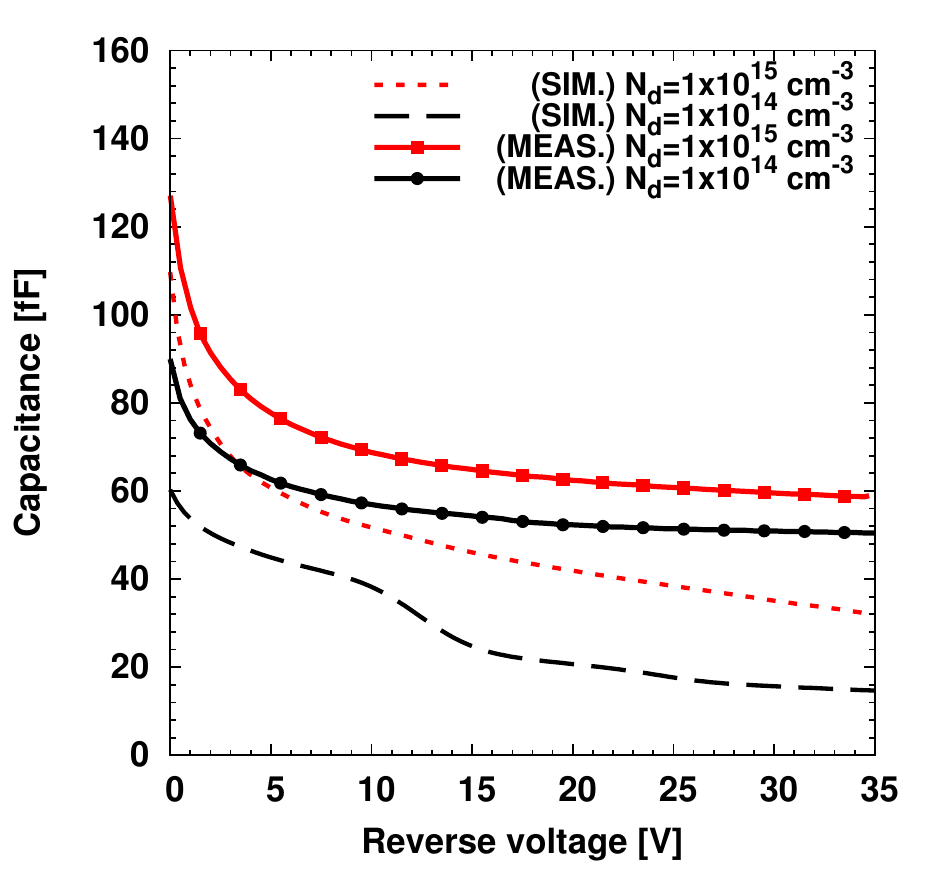}
}%
\caption{Comparison between measured and simulated electrical quantities for sensors having two different 
bulk doping concentrations.} 
\label{fig7}
\end{figure}
\subsection{Current-Voltage (I-V) measurements}
The measured current-voltage (I-V) curves for two fabricated devices of two different resistivities 
(10~$\Omega$cm and 100~$\Omega$cm) together with the obtained results from 3D numerical simulations
are shown in Figure~\ref{fig7a}.
As expected, the breakdown voltages are low, between 30 and 40~V, 
with the sensor of higher doping concentration showing earlier breakdown.
The absolute current levels are low, typically less than 2.5 pA per strip. 
The leakage currents therefore have a low noise contribution during sensor operation. 
The numerical simulations (dashed lines) are in good agreement with the measurements in terms of both
current values and breakdown voltages. The small discrepancies are mainly related to three effects:  
(i) the dependence of current levels on local carrier lifetimes that can vary from sensor to sensor, 
(ii) the dependence of breakdown voltages on uncertainties in the p-spray profile, and (iii) the refinement 
of the numerical grid which is limited by the available computational power.

\subsection{Capacitance-Voltage (C-V) measurements}
Results from capacitance-voltage (C-V) measurements and simulations are shown in Figure~\ref{fig7b}. 
The measured C-V curves for both sensor types decrease at a similar rate for low applied voltages. 
The curves then tend to saturate as the reverse voltage is increased.
A full capacitance saturation cannot be reached before the breakdown of both sensors. This is caused by the large 
contribution of the n$^+$ to p-spray junction capacitance along the strip perimeter. This region is depleted
at a much slower rate and is never fully depleted, thus the absence of saturation.
In addition to this effect, full depletion cannot be reached for the device with the higher doping concentration,
causing its capacitance to be slightly larger than that of the other device 
(it is estimated, from theoretical calculations, that this device requires roughly 75V to be fully depleted). 
Given the obtained average capacitance value of $\sim$65~fF per strip, the sensors contribution to the noise 
of the readout system should be very low.
The simulated C-V curves are in agreement with the measured ones apart from a few details.
The slope variation between 5 and 15~V for the simulated device of lower bulk doping cannot be 
observed in the measurements because of other parasitic effects in the measurement setup. 
In addition, the simulated capacitance values present a negative offset with respect to the measured ones. 
This shift can be attributed to parasitic capacitances and MOS effects underneath the long metal lines of 
the fabricated sensor that cannot be reproduced in the simulations due to the insufficient amount of memory
available for the computations.

\section{Functional characterisation}\label{sec:functional-charact}
Functional testing on the fabricated detectors was carried out at the ESRF at two different beam lines.
The operation of the strip sensors in MRT conditions was examined at ID17 \cite{ID17} using a multi-channel 
readout system to verify their detection capability of full intensity micro-beam array.
In addition, another set of tests were performed using an X-Ray scanning microscope at 
ID21 \cite{ID21} to observe the signal efficiency of the strips with bi-dimensional scanning using low energy X-rays.
The results reported here focus on the most relevant layout configurations. Table \ref{tab2}
summarises the properties of the devices under test. All sensors are given an identifier (ID), according
to wafer number, strip length, type of steering ring and a number corresponding to the location of the sensor 
on the wafer.
For example, sensor \#1 is from wafer A7, has a strip length of 100~$\mu$m, a closed steering ring and was 
number 14 on the wafer.
\begin{table}[t]
\caption{Summary of the characteristics of the sensors object of this study.}
\vspace{0.25cm}
\begin{center}
\begin{tabular}{lccccc}
\toprule
\textbf{\#} & \textbf{Sensor ID} & \textbf{Bulk doping} & \textbf{Strip Length} & \textbf{Layout type} & \textbf{channels}\\
- & - & \textbf{[at.~B/cm$\mathbf{^3}$]} & \textbf{[$\mathbf\mu$m]} & - & - \\
\midrule
1 & A7-L100-CL-14 	& 10$^{15}$	& 100	& closed-ring & 128\\
2 & A7-L250-CL-13 	& 10$^{15}$	& 250	& closed-ring & 128\\
3 & A2-L050-CL-15 	& 10$^{15}$	& 50		& closed-ring & 128\\
4 & A9-L050-VAR-03	& 10$^{14}$	& 50		& closed-ring + variable pitch & 128\\
\bottomrule
\end{tabular}
\end{center}
\label{tab2}
\end{table}%

\subsection{Beam-tests at ESRF ID17}\label{subsec:Beam-tests at ESRF ID17}
\begin{figure}[!t]%
\centering 
\subfigure[]{\label{fig8a} 
	\includegraphics[width=.45\textwidth]{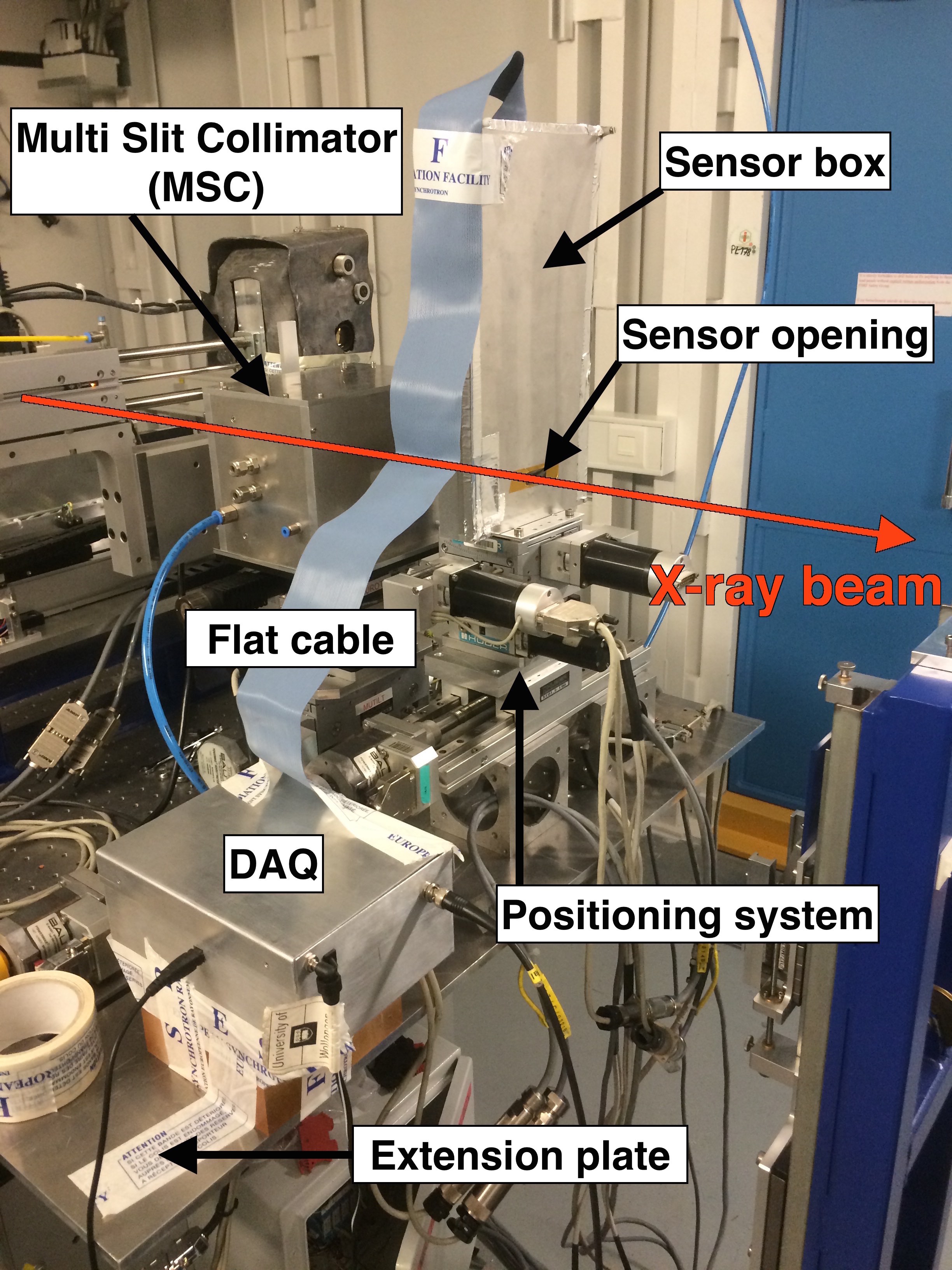}
}%
\subfigure[]{\label{fig8b} 
	\includegraphics[width=.45\textwidth]{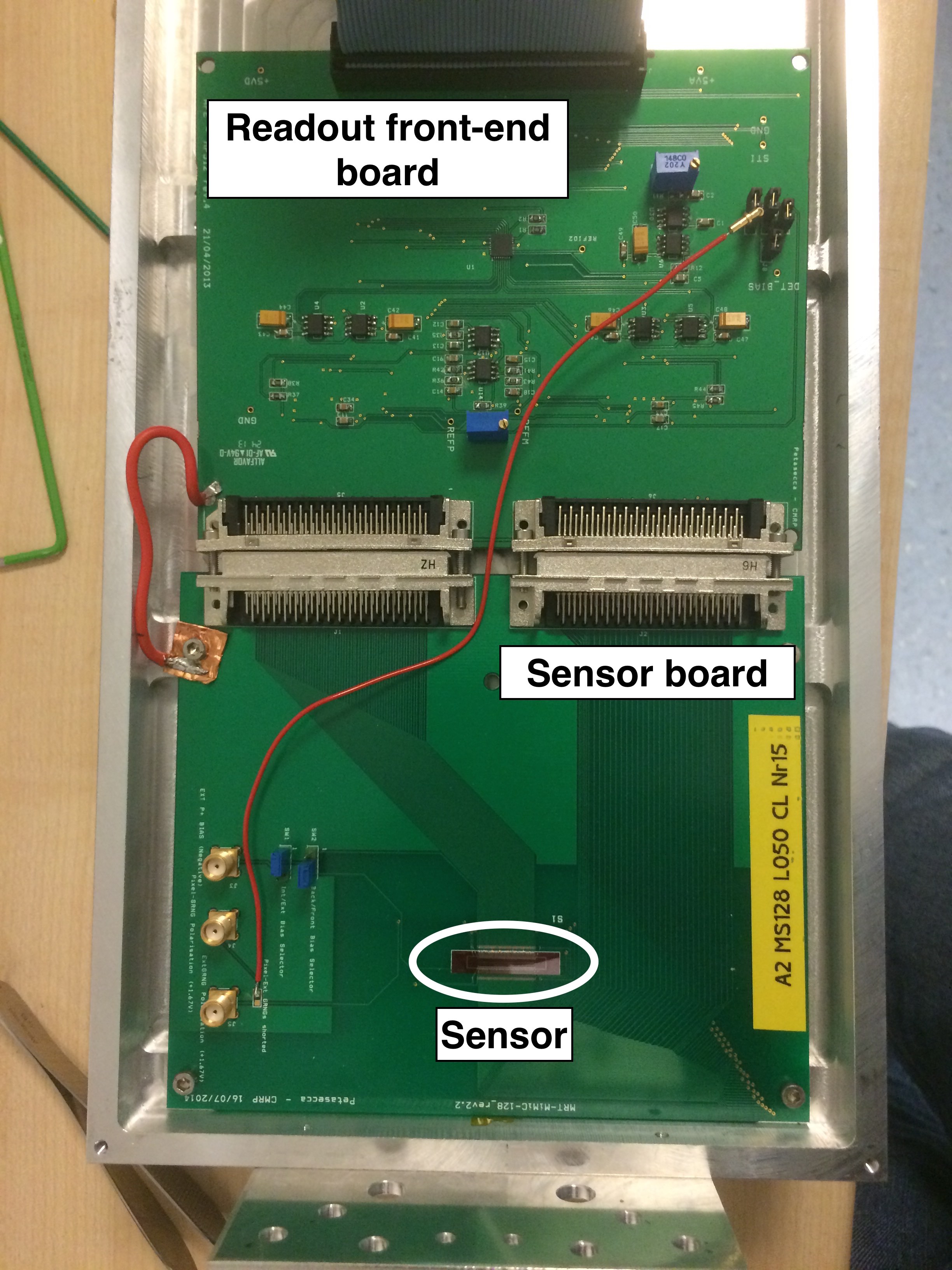}
}%
\caption{Image of the measurement setup installed at ID17. External view of the entire system (a) and 
details of the inside of the shielding box (b) are shown.} 
\label{fig8}
\end{figure}
The ID17 beamline is used for ongoing pet trials. The planned experiment had to be adapted to suit the 
specific environment.
All the required equipment for the delivery of the MRT beam in its fractionated configuration is mounted on an
optical table. The positioning system required for the testing of the beam monitors is comprised of 3 motorised
stages that were aligned to the main axis of the beamline. A metal shielding box containing the sensor and the first 
stage of the readout electronics is placed on top of the positioning system (Figure~\ref{fig8a}).
The metal box is required in order to shield the system from electromagnetic noise present in the experimental 
hatch. The sensor to be tested was glued and wire bonded to a printed circuit board (PCB) that was connected 
to the multichannel data acquisition system (DAQ). 

The electronics used in these tests is based on the Texas Instruments AFE0064 chip \cite{AFE0064}, 
a 64-channel analog front end for X-ray detectors. The chip includes an FPGA for gain level selection, 
a correlated double sampler, a 64-channel multiplexer, and two differential output drivers. 
The sensor is readout in a continuous mode. The integration time and channel range can be manually selected. 
The signals are sent to an FPGA which then sends the information to a PC via a USB connection. 
Figure~\ref{fig8b} shows the sensor wire-bonded to the support PCB which is connected to the readout front end. 
The connection to the FPGA in the DAQ system  is performed through a flat cable visible in Figure~\ref{fig8a}. 
The sensor positioning and scanning across the beam is performed using the SPEC\textsuperscript{\textcopyright}  
control software \cite{SPEC} through a computer located in the control room. 
The minimum step size of the motors used was 10~$\mu$m. 
The two beam tests at ID17 were performed under  different beam conditions, 16-bunch mode
and uniform mode with the latter delivering higher radiation doses. 

\subsubsection{Single strip measurements}
Initial tests were performed with the synchrotron running in 16-bunch mode, in order to start the characterisation
under less challenging conditions (lower delivered doses). The characterisation of a single strip would allow 
simple verification as well as quick understanding of the sensor performances under the operation conditions 
of ID17. 
In this measurement, the central strip (strip number 64 on a 128 channels sensor) and the steering ring were 
connected to two separate electrometers. The ID of the sensor used was A7-L100-CL-14 (see Table~\ref{tab2}), 
and the measurements were performed in passive mode (without bias).
The sensor was scanned across an array of 25 micro-beams having a
peak-to-peak (centre-to-centre) distance of 400~$\mu$m, with the single X-ray micro-beam width set at 
50~$\mu$m.
\begin{figure}[!t]%
\centering 
\subfigure[]{\label{fig9a} 
	\includegraphics[width=.33\textwidth]{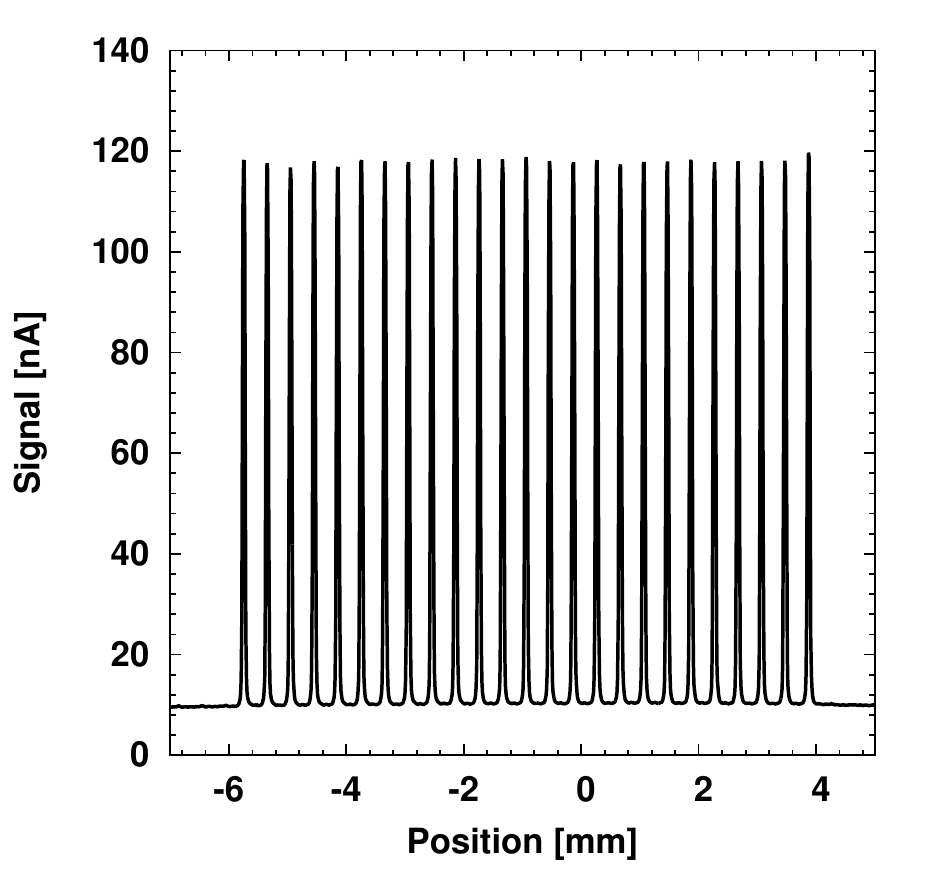}
}%
\subfigure[]{\label{fig9b} 
	\includegraphics[width=.33\textwidth]{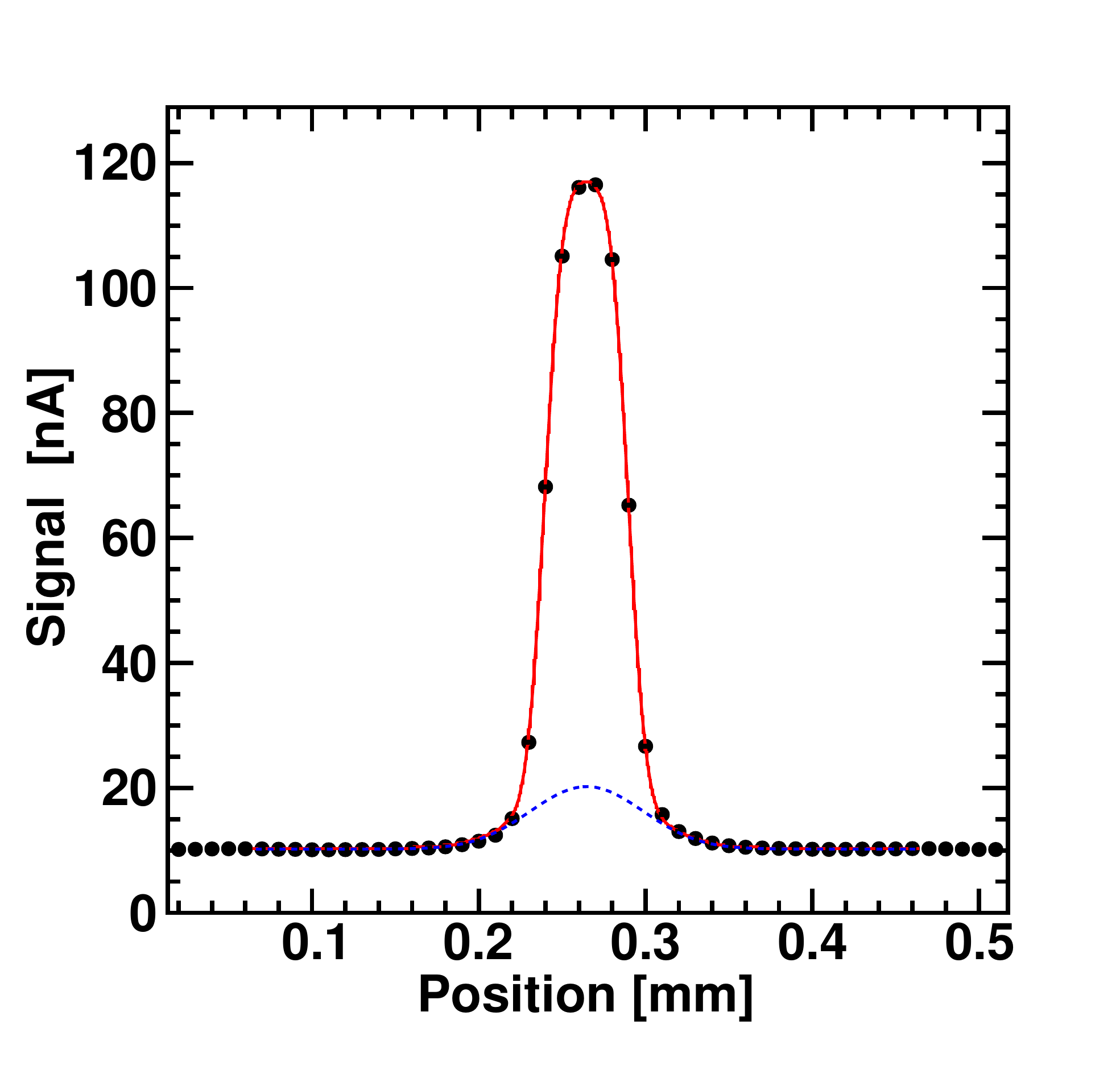}
}%
\subfigure[]{\label{fig9c} 
	\includegraphics[width=.33\textwidth]{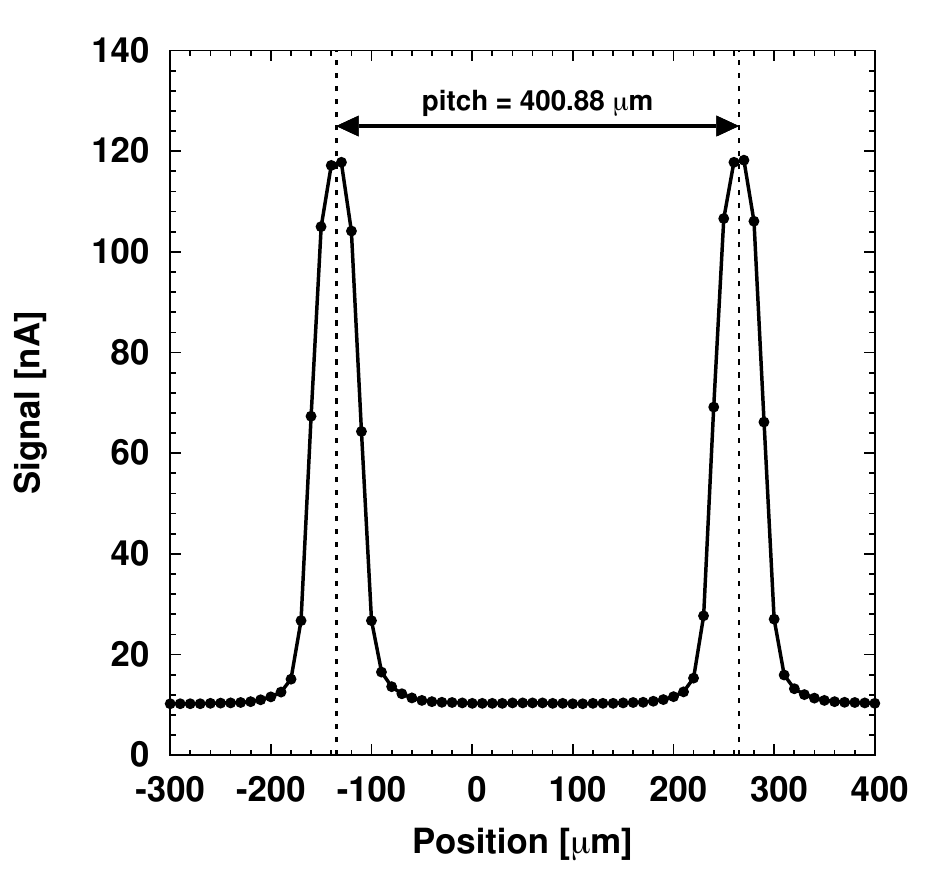}
}%
\caption{Single strip measurement results for a sensor with closed steering-ring and strip length of 100~$\mu$m
operated in passive mode (0V). Here reported are the measured profile for an array composed by 25 
micro-beams (a), the measured shape of a single micro-beam and the expectation from the fitted model (see text) 
(b) and the measured peak to peak pitch (c).} 
\label{fig9}
\end{figure}
The gap of the beamline's Wiggler source was set to 24.8~mm, delivering the highest dose rates. 
The scanning step was 10~$\mu$m and the electrometer currents were recorded at each scanned position 
as shown in Figure~\ref{fig9}.
The profile of the micro-beam array was successfully reconstructed by the device under test and the high 
dose regions (or peaks) were clearly measured and detected. 
A good amplitude uniformity was found, with a current value 
of roughly 120~nA per peak. The small current value indicated that the steering ring serves its purpose 
to limit the current generated in adjacent strips. The smearing of the micro-beams is minimal and near zero
signal is detected in the valley regions.
The shape of a single micro-beam can be observed in Figure~\ref{fig9b}. The measured values are shown with 
dots while the fitting model is shown with the red line. 

A model for fitting the shape of the peak can be devised by assuming the ideal intensity profile
to be a square function ("box") with full width given by the slit width (nominally 50 microns in this case). 
The two causes for the measurement deviations from this model are: (i) the silicon sensor has a smeared charge 
collection profile (as shown in Figure~\ref{fig5}) and (ii) the physics processes (beam divergence, scattering etc.)
that the X-rays are subjected to in their path from the collimator entrance to the sensor. 
The edges of the "box" profile thus appear to be smeared.
A first attempt to improve the description of the edges was done by convoluting the box
with a gaussian. A fit to such a model gave a good description of the intensity profile in the peak area, but did 
not account properly for the tails. An additional box of the same total width, smeared with a second 
gaussian was added to the model achieving a much better representation of the tails of the measured micro-beam.
The blue dashed line in Figure~\ref{fig9b} represents the contribution of this second profile, while the solid red 
line shows expectation from the full fitted model.
This model was used to separately fit each of the 25 peaks in the measured micro-beam array. 
Using the r.m.s. spread of the fitted results as error margin the following results are obtained. 
The full width of the initial box profile was found to be 49.4$\pm$0.8~$\mu$m. 
The sigma of the first gaussian was 8.3$\pm$0.3~$\mu$m. 
This is consistent the shape of the charge collection efficiency profile of the sensor
(Figure~\ref{fig5}). The sigma of the second smeared box was 29.0$\pm$2~$\mu$m, contributing to roughly 10 to 
15\% of the peak.  The FWHM of the final fitted profile was found to be equal to 52.0$\pm$3~$\mu$m,
consistent with the expectation for a 50~$\mu$m wide micro-beam.
The pitch between two adjacent micro-beams at the measurement point 
(roughly 15 cm from the MSC) was estimated to be 400.9~$\mu$m (Figure~\ref{fig9c}).
This estimation suggested that beam divergence is present at the point of measurement and alignment 
difficulties when using the multichannel system are to be expected.
 
\begin{figure}[!t]%
\centering 
\subfigure[]{\label{fig10a} 
	\includegraphics[width=.45\textwidth]{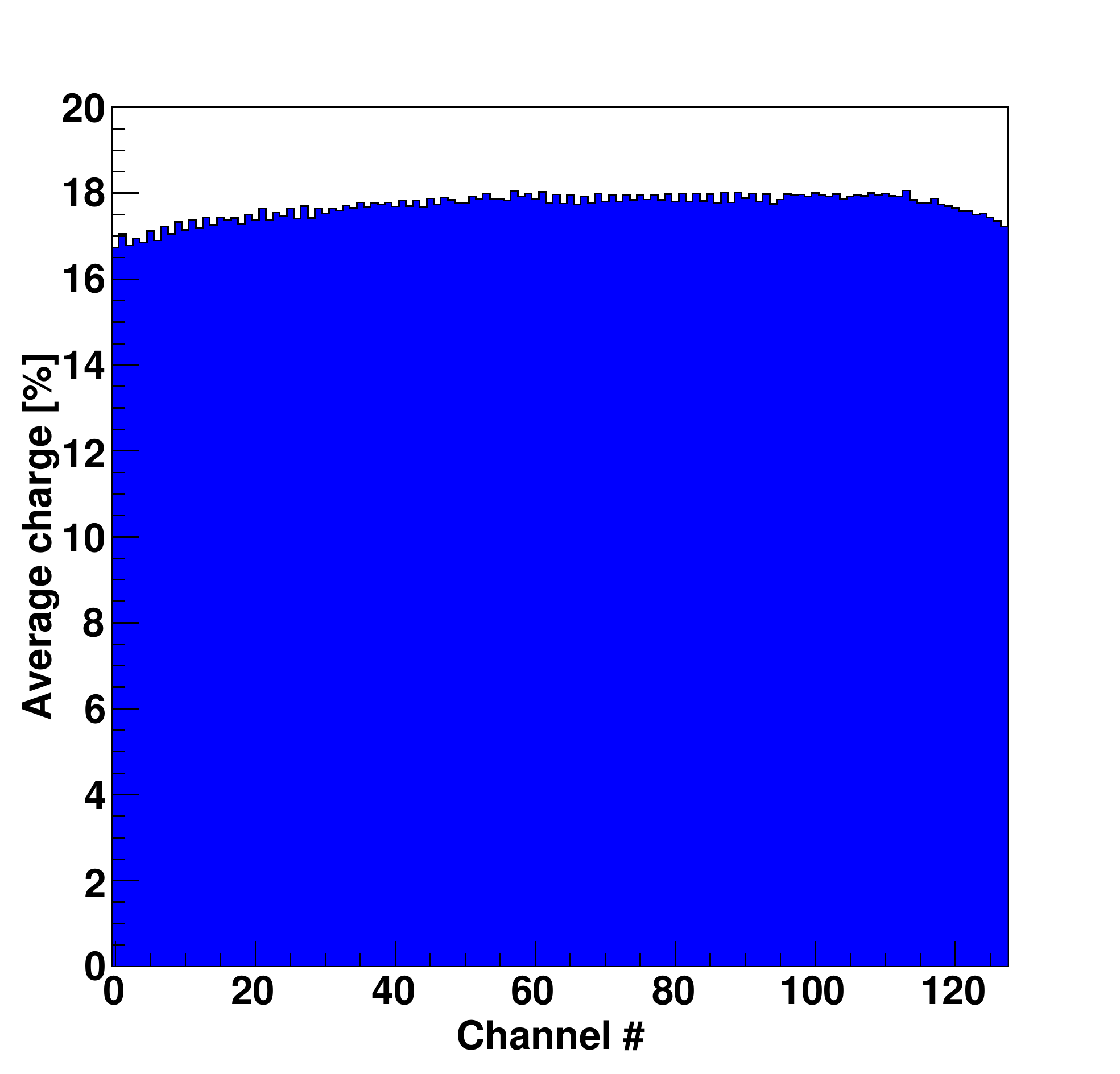}
}%
\subfigure[]{\label{fig10b} 
	\includegraphics[width=.45\textwidth]{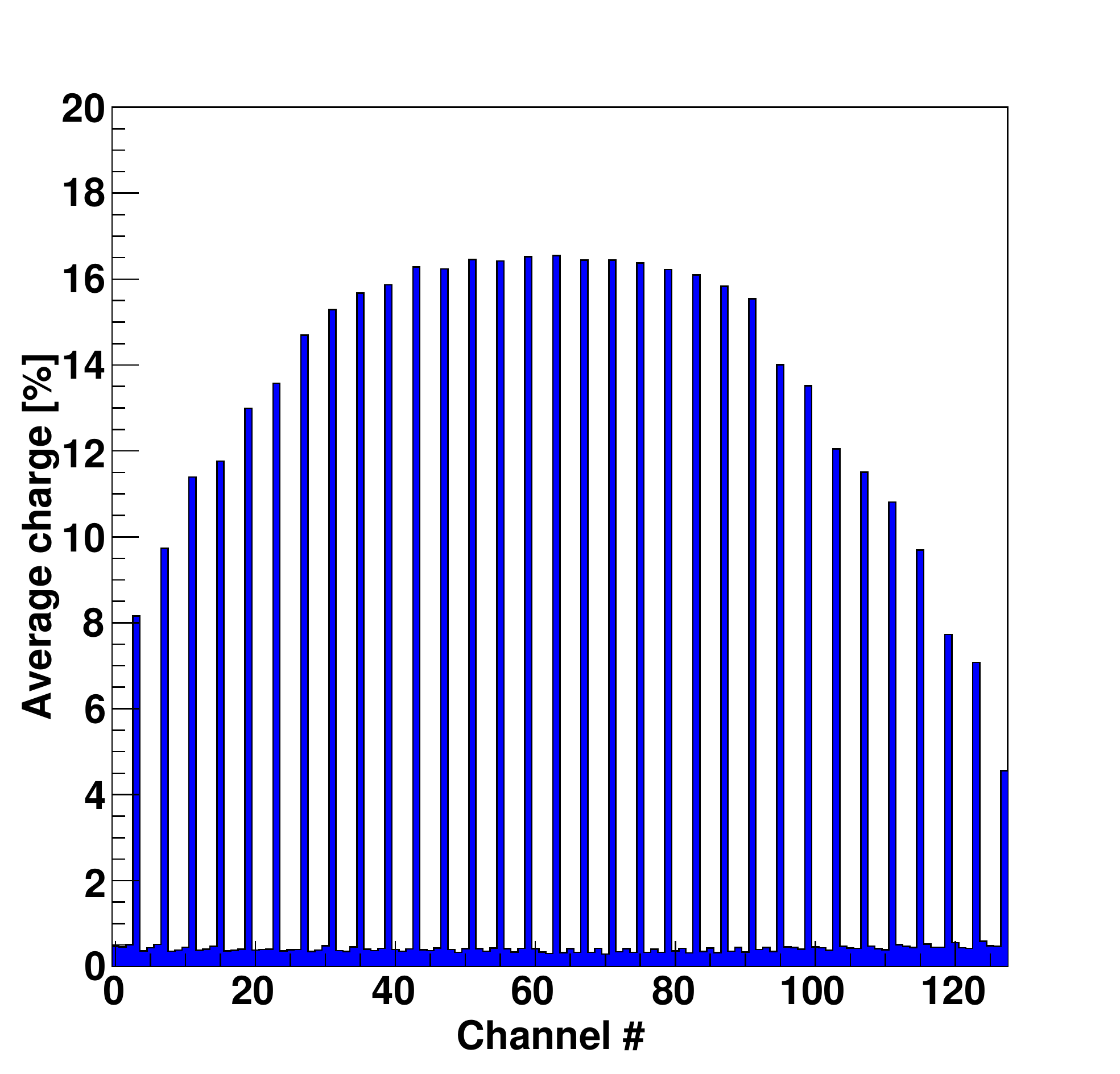}
}%
\caption{Measurements performed with the multichannel DAQ and the synchrotron running in 16-bunch mode
(sensor ID A7-L250-CL-13). Shown here are a broad beam measurement (a) and an
acquired MRT beam profile (b).} 
\label{fig10}
\end{figure}
\subsubsection{Multi-strip measurements}
The second test focused on the calibration and the use of the digital DAQ system. 
This test was first performed on sensors with the longest strip length of 250~$\mu$m (sensor ID A7-L250-CL-13 in
Table~\ref{tab2}).
The basic settings of the readout electronics, including the integration time and full scale current range for 
each channel were identified using measurements from broad beam mode.
A result obtained in broad beam configuration is shown in Figure~\ref{fig10a}. 
With the wiggler gap set to 24.8~mm, the beam was collimated to 13.7~mm wide and 
0.102~mm high, covering the entire active area of the device under test. 
The results are reported as percentage of the maximum charge 
measurable per channel. A maximum value of $\sim$18~\% was obtained, which is still far from the saturation
of the electronics. In addition, this measurement verified that all strips on the sensors were functional, 
with no defective or noisy channels. 

The second test was performed by illuminating the sensor with an array of micro-beams where its capability 
to record the micro-beam array was assessed.
The array of micro-beams had a pitch of 400~$\mu$m, four times of the inter-strip 
pitch of the sensor design. As a result, if the first channel monitors a peak, the second peak will be detected 
in the fifth channel leaving three channels to monitor the valleys.
The sensor was manually aligned such that the central strip was centred with respect to the micro-beam array. 
The wiggler gap was set to 24.8~mm, and the sensor was then exposed to an array of 33 micro-beams for 
approximately 2 seconds. The data were recorded by the sensor in passive mode. 
Each micro-beam was 50~$\mu$m wide 
and 102~$\mu$m high. The total width of the array was measured to be 13.2~mm, covering the entire active 
area of the sensor which is 12.7 mm wide. 
The key observations obtained from the results in Figure~\ref{fig10b} are as follow: (i) the three channels between 
each channel where a peak was recorded have near zero signal or are "off" as expected from the known profile of 
the micro-beam array; (ii) no cross-talk was observed between adjacent channels, indicating that the 
implementation of steering ring limits cross talk as expected;
(iii) only $\sim$18\% of the full dynamic range of the electronics was used, which implies that saturation in 
the readout system has been totally eliminated using the particular beam intensity in this experiment; 
(iv) a significant decrease in intensity was observed on both sides of the overall profile measured at $\sim$15~cm
from the MSC.
The observed intensity decrease is caused by the beam divergence at the measurement position. 
As shown in Figure~\ref{fig9c}, the pitch between two micro-beams was measured to be 400.9~$\mu$m 
instead of 400.0~$\mu$m at the MSC.
This resulted in a cumulative misalignment between the sensor and the micro-beams on the 
periphery of the device. In fact, as demonstrated by the
numerical simulations in Figure~\ref{fig5}, the strip active volume has FWHM of $\sim$17~$\mu$m, so it can be
estimated that the strip signal will start to decrease significantly from the 9$^{th}$ channel from the central strip.
This is in agreement with Figure~\ref{fig10b}, where the response of approximately 17 channels located in the 
central region of the sensor is relatively flat. The signals gradually decreases with each channel that is further 
away from this central region. 
This effect can be minimised by more detailed alignment of the sensor with respect to the MSC which is rather 
limited with the current beam-line configuration at the ID17.
Possible solutions are: (i) re-design sensors that have an inter-strip pitch that matches the pitch of the MRT 
array at a specific position, (ii) producing a sensor with a slightly larger strip pitch and then acting on the sensor 
tilt in the "z" plane to match the pitch of the MRT micro-beams (more flexible) or (iii) producing a new MSC in 
which the sensor can be mounted just a few millimetres downstream of the generation point of the micro-beam 
array, thus minimising the effect of beam divergence at the measurements position. 
Both options one and two provide little flexibility for the system. The last solution would be the most optimal from 
the measurement point of view, but can be costly and complex to implement.

\begin{figure}[!t]%
\centering 
\subfigure[]{\label{fig11a} 
	\includegraphics[width=.33\textwidth]{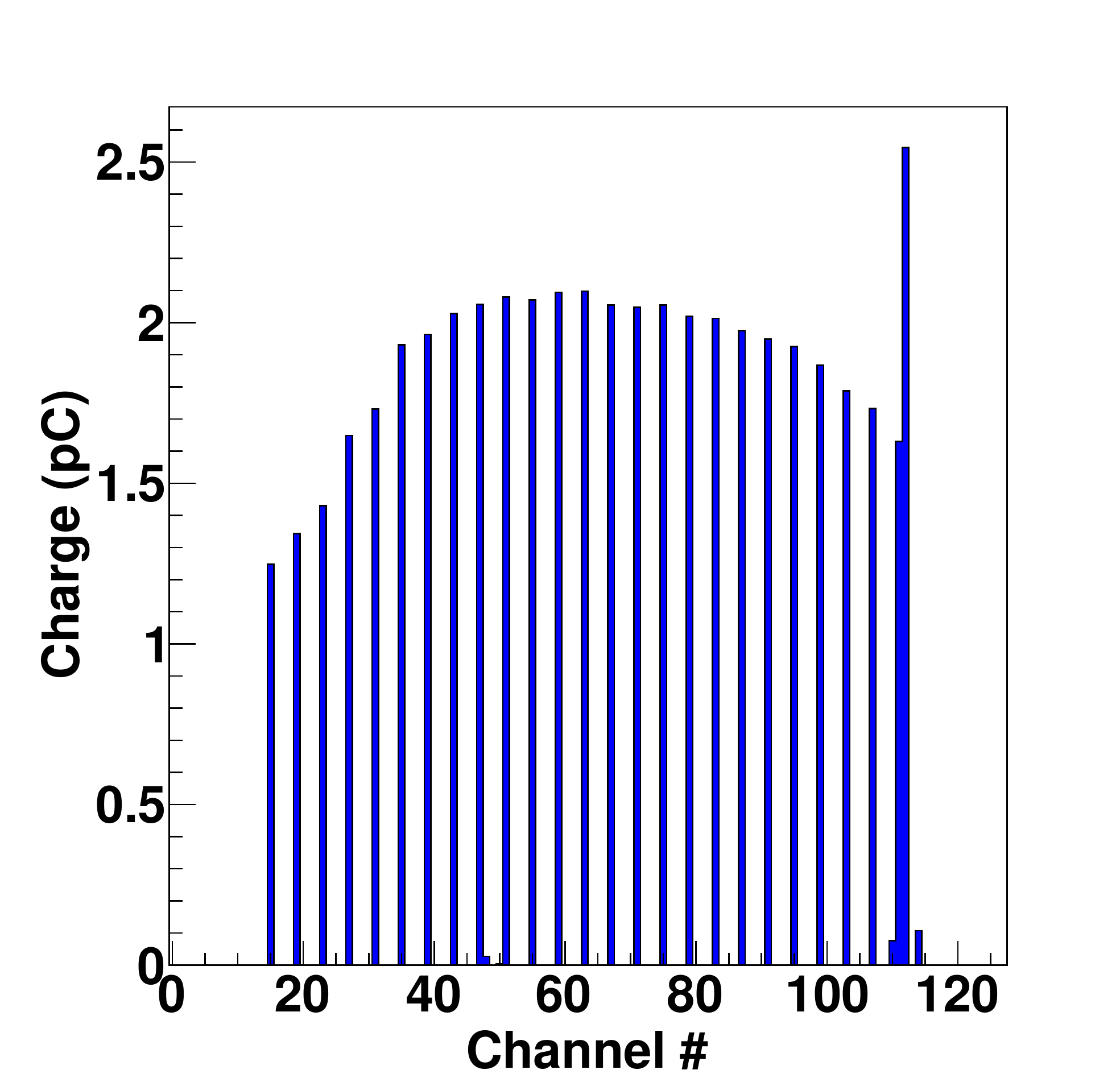}
}%
\subfigure[]{\label{fig11b} 
	\includegraphics[width=.33\textwidth]{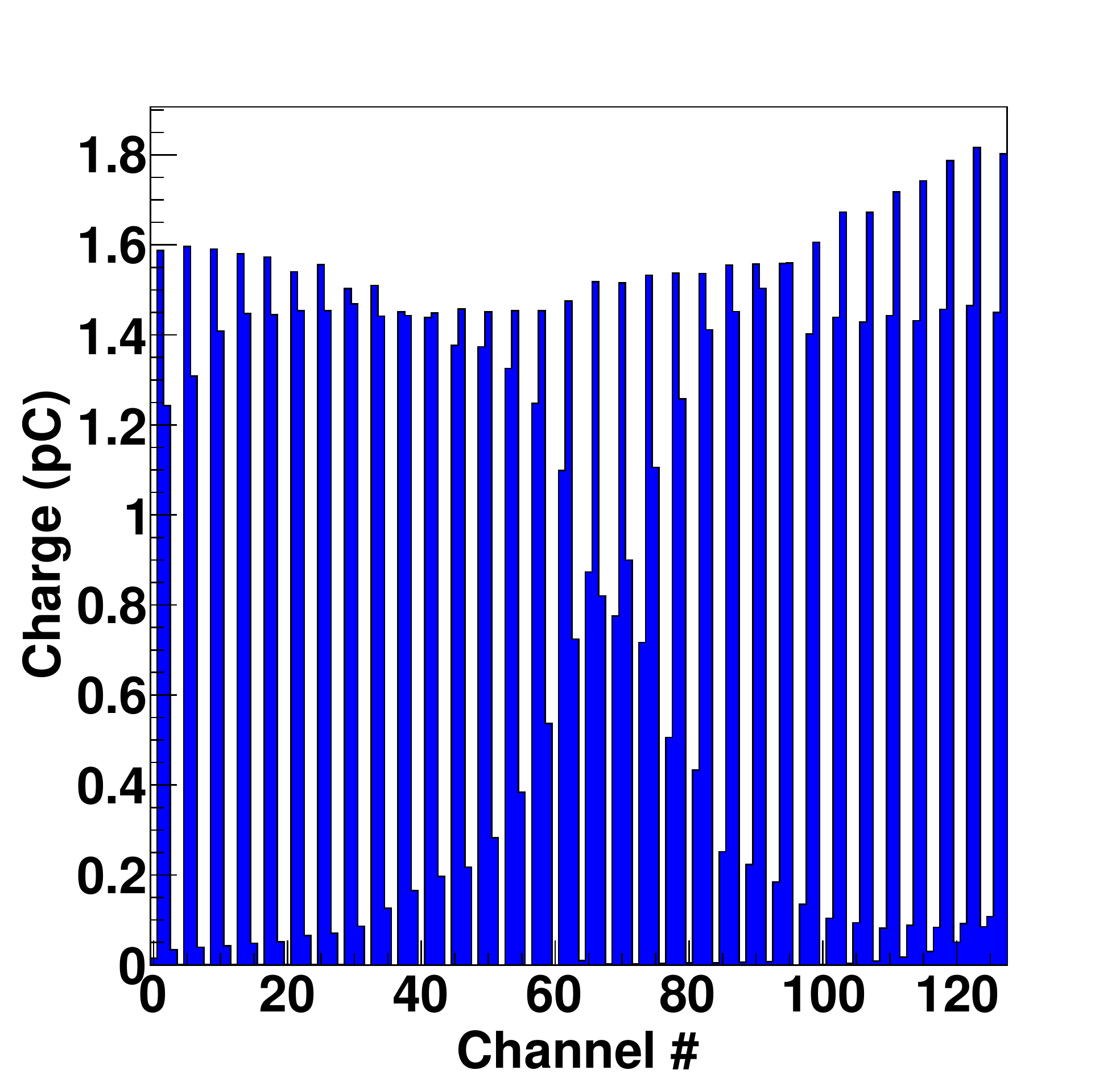}
}%
\subfigure[]{\label{fig11c} 
	\includegraphics[width=.33\textwidth]{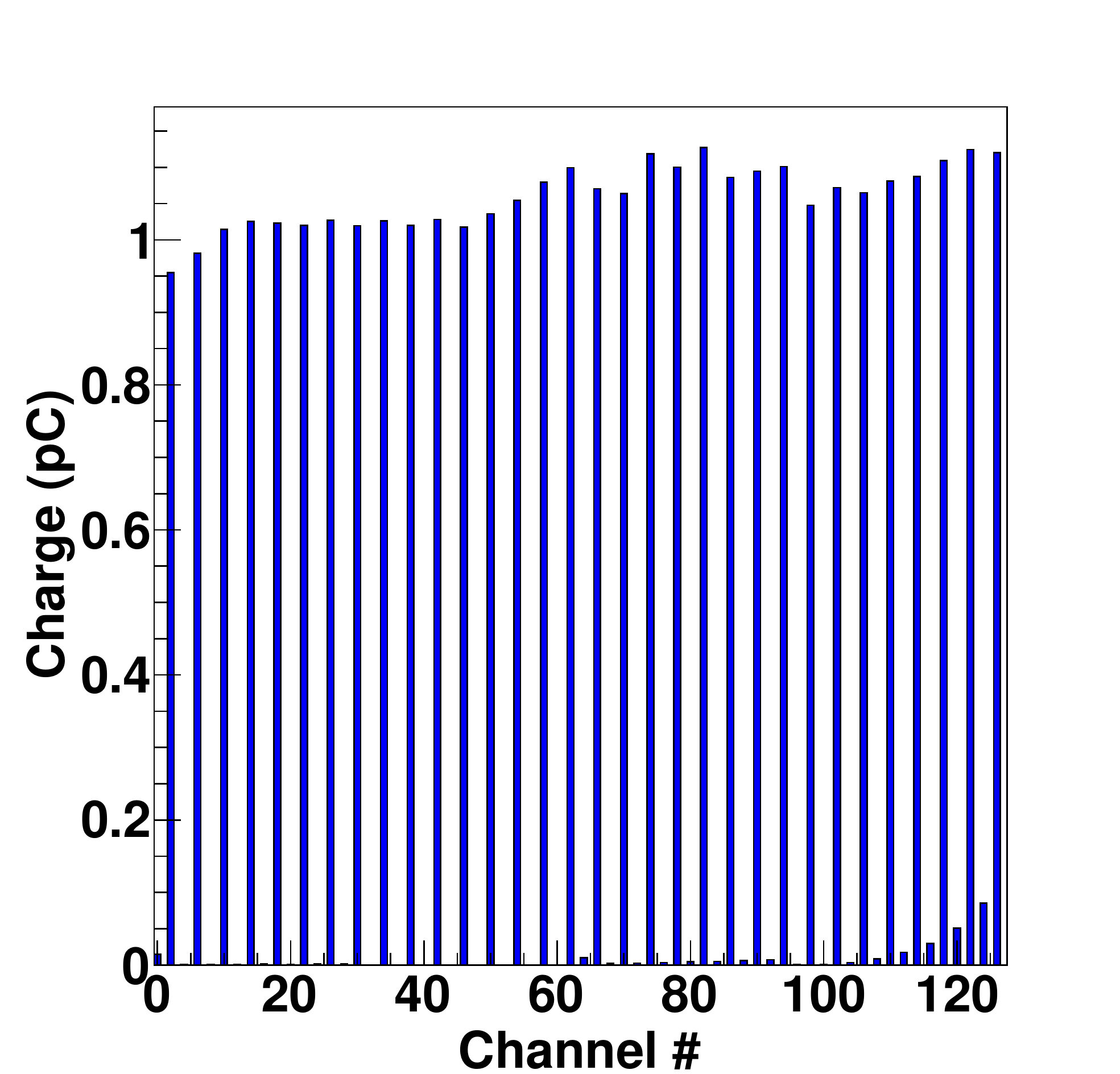}
}%
\caption{Measurements performed on two different sensors with a full intensity MRT beam. 
Standard sensor (A2-L050-CL15) (a) and variable pitch sensor (A9-L050-VAR03) (b) and the profile
obtained by averaging the signals of the triples (c).} 
\label{fig11}
\end{figure}

The sensors were also tested with the synchrotron operating in uniform mode delivering the full beam intensity 
at the ID17. From the experience gained in the previous test, the sensor was moved as
close as possible to the MSC while maintaining a simple mounting scheme.
A new shielding box and mounting were put in place, allowing the MSC to sensor distance to be decreased
to roughly 10~cm. Because of the much higher intensity a sensor with a shorter strip
length (50~$\mu$m) was used to reduce the intensity of the acquired signal (sensor ID A2-L050-CL-15). 
In addition, a sensor with a variable strip pitch (Figure~\ref{fig1c}) was also tested, aiming at a more precise 
peak measurement (3 points per peak instead of 1) and to counteract the beam divergence
(sensor ID A9-L050-VAR-03). Measurements were performed in passive mode and with beam arrays of 25 and 
33 peaks. The wiggler gap was set to 24.8~mm. The measurement results are shown in Figure~\ref{fig11}.
For the standard sensor (Figure~\ref{fig11a}) a similar pattern as in previous measurements was observed
(compare with Figure~\ref{fig10b}), although in this case the lateral decrease was less pronounced due to the
reduced distance to the MSC. In this case it is also possible to observe a noisy channel, probably a defective strip,
at channel 110. The measurement obtained with the variable pitch device (Figure~\ref{fig11b})
suggests that, with this sensor geometry, it might be possible to effectively 
counteract the beam divergence even in sub-optimal operating conditions. The expected result for a 
strip "triplet", is to have a large signal on the central strip and lower signals on the two neighboring
channels. In Figure~\ref{fig11b} this occurs on strip 66, also suggesting that the beam is correctly aligned 
on this channel. Proceeding to the sides of the measured profile, the outermost strip of the "triplet" sees a
gradual signal increase while the innermost one sees a comparable decrease until no signal is visible 
anymore. At this point, it is possible to calculate the arithmetic average of the signals of the triplets and assign
the results to the central strip, obtaining a rather flat profile as reported in Figure~\ref{fig11c}.
This solution, combined with future sensor and setup modifications should allow a reliable and consistent 
monitoring of the MRT beam. 
Considering the signal level of the two different devices examined here, the difference in results from the two 
different bulk resistivities were insignificant. High resistivity materials are the most widely used material because 
of their high purity. Moreover, their widespread use in silicon radiation detector fabrication, provides that their 
properties and behaviours are much better understood than the use of low resistivity materials. 
For future development of sensors for the MRT, high resistivity materials would be a much preferred choice.  

The final study in this experiment concerns signals in the valley regions. It is known that some residual radiation 
will be present in the valley regions. Because of the large difference in magnitude between
the peaks and the valleys, it was not possible to find a readout setting that allowed
the simultaneous acquisition of the signal in both positions. 
This is a crucial setting when the system would be used to monitor the array of micro-beams precisely.
This problem can be solved by separately tuning the dynamic ranges for each individual channel corresponding 
to the peak and the valley regions using a new version of the control software for the DAQ system.

\subsection{Beam-test with sub-micron X-ray beam at ESRF ID21}
Using the X-ray scanning microscope available at beam line ID21 at ESRF, accurate 2D scans of the 
devices under test were performed. The signal response to a sub-micron X-ray beam of 7.2~keV was recorded
as a function of the hit position. The focused beam size was 
$\sim$0.35$\times$0.94~$\mu$m$^2$. 

\begin{figure}[!t]%
\centering 
\subfigure[]{\label{fig12a} 
	\includegraphics[width=.35\textwidth]{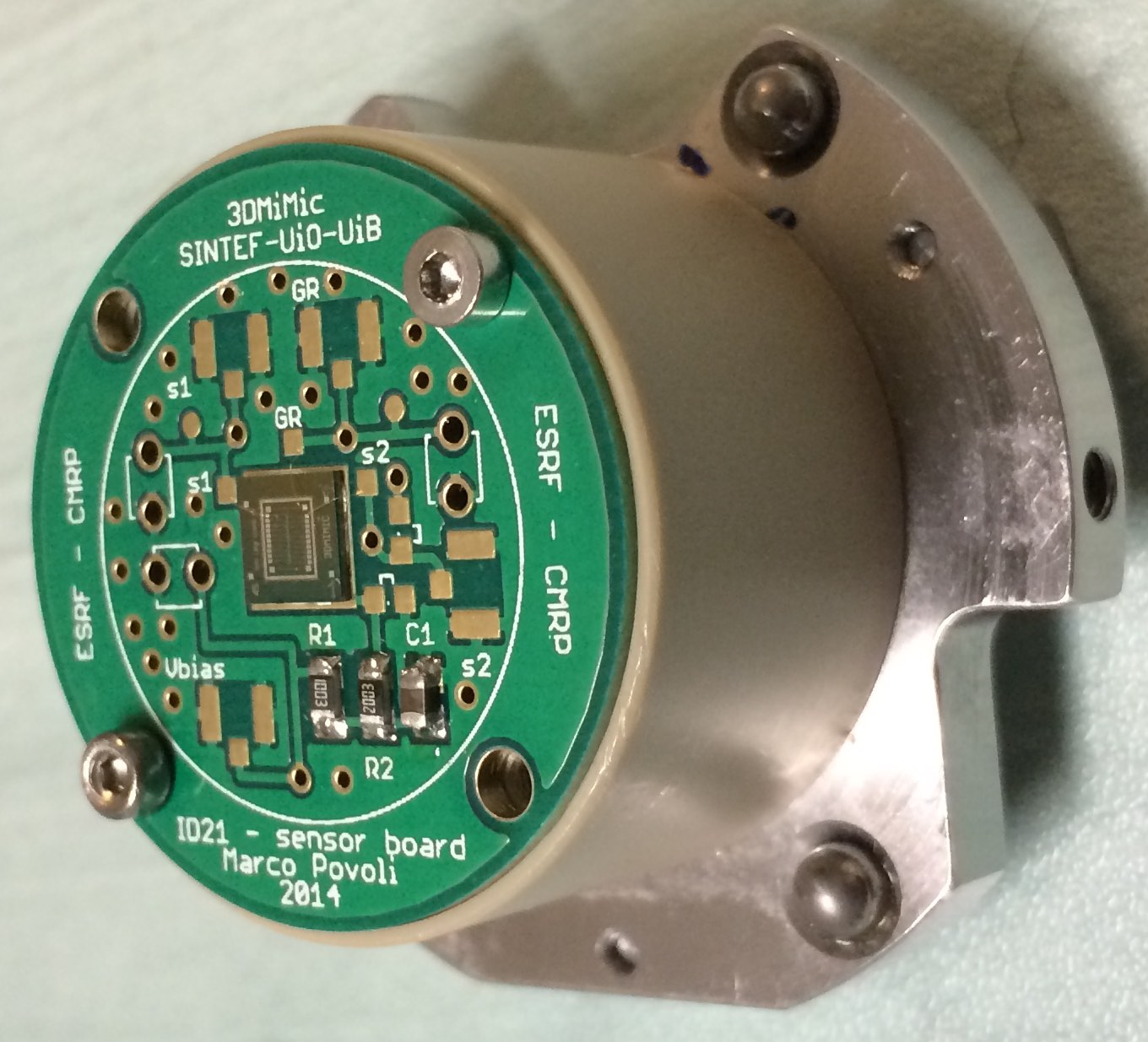}
}%
\subfigure[]{\label{fig12b} 
	\includegraphics[width=.24\textwidth]{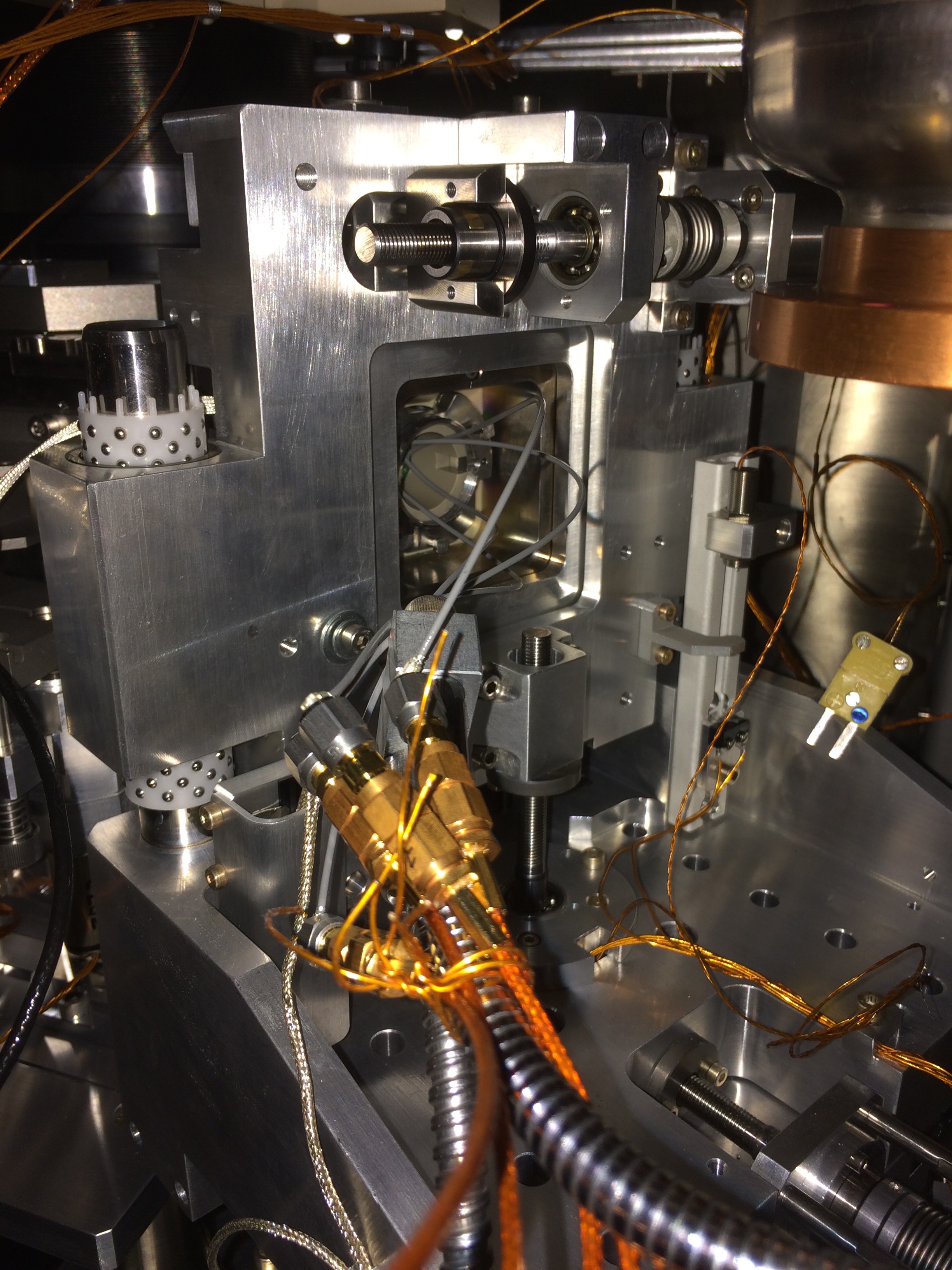}
}%
\subfigure[]{\label{fig12c}
	\includegraphics[width=.37\textwidth]{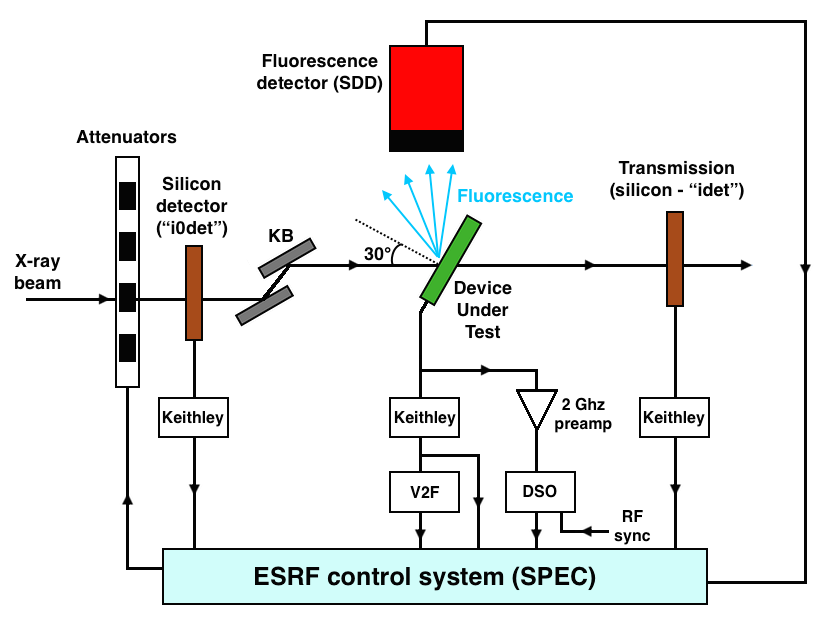}
}%
\caption{Image of a sensor mounted on the specific ID21 sample holder (a), picture of the inside of the 
SXM vacuum chamber (b) and simplified schematics of the beam line (c).} 
\label{fig12}
\end{figure}

The devices under test consist of a series of single strip elements the same as those in the multi-strip sensors 
tested at the ID17. In order to avoid the complexity of a multi-channel DAQ, several single elements were 
connected by a single metal line to allow single channel readout.
The sensors were mounted on a PCB designed to fit on the standard ID21 sample support to be used 
during testing. An example of a sensor and PCB assembly mounted on the support is shown in Figure~\ref{fig12a}.
The sample holder is inserted in the back of the scanning X-ray microscope (SXM, Figure~\ref{fig12b}) 
and the connections were made using micro-coaxial cables that are connected to vacuum compliant SMA 
cables. The electrical connections are fed out of the vacuum chamber through a flange. 

A schematics of the the ID21 beamline is shown in Figure~\ref{fig12c}. A set of attenuators 
can be inserted in the beam to modulate its intensity. 
The beam intensity variation was constantly monitored both upstream and downstream of the device under 
test by two separate silicon detectors, allowing the user to perform any final correction on the acquired data.
The X-ray beam hits the investigated sample at an angle of 60\textdegree to its surface. 
In addition, a silicon drift detector (SDD) can be used to observe the X-ray fluorescence 
coming from the device under test. Two separate readout configurations were available: (i) the currents of the 
strip and the steering-ring were recorded using two Keithley electrometers and plotted, (ii) the evolution of the 
sensor signal as a function of time was studied using a 2.3~GHz bandwidth preamplifier and a Digital Sampling 
Oscilloscope (DSO) with 3~GHz bandwidth. 
Scanning of the device under test and data acquisition were automated using the  
control software available at the ID21 beamline.

\begin{figure}[!b]%
\centering 
\subfigure[]{\label{fig13a} 
	\includegraphics[width=.32\textwidth]{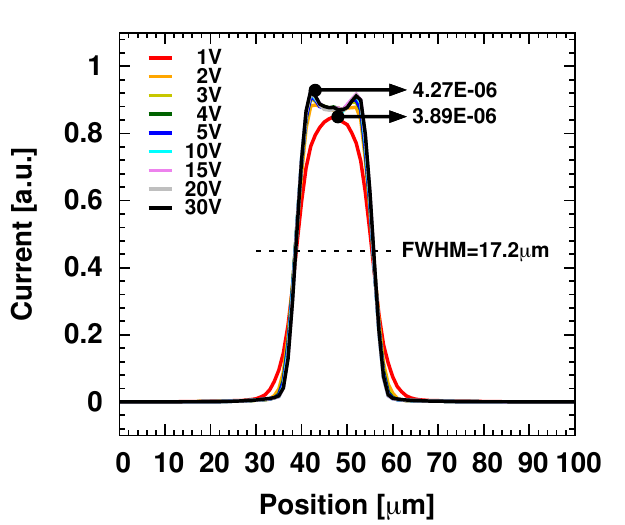}
}%
\subfigure[]{\label{fig13b} 
	\includegraphics[width=.32\textwidth]{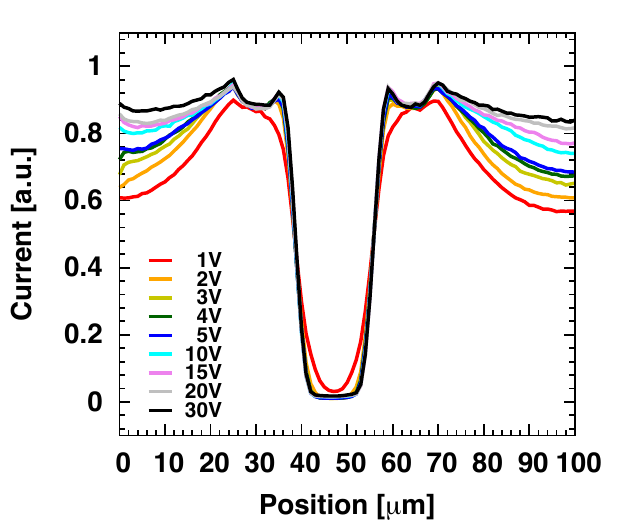}
}%
\subfigure[]{\label{fig13c}
	\includegraphics[width=.32\textwidth]{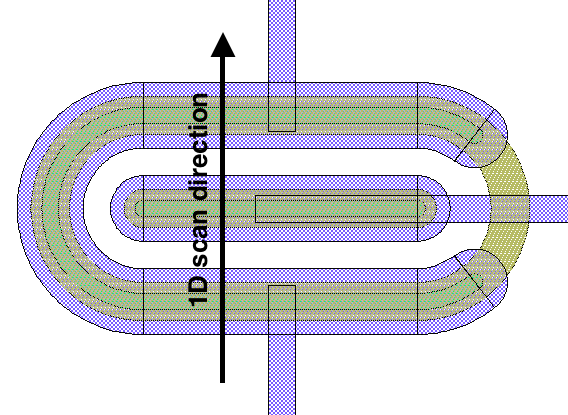}
}%
\caption{Simple 1D scan performed on a sensor from wafer A9 with a strip length of 50~$\mu$m and a closed
steering-ring. Strip signal (a) ring signal (b) and layout with scan direction (c) are shown.} 
\label{fig13}
\end{figure}

\subsubsection{Signal response scans using Keithley electrometers}
The measurements were performed in 1D (simple line scan) or in 2D scans (longer duration). 
The results reported here are for a sensor from wafer A9 (1$\times$10$^{14}$~at.~B/cm$^3$) with a 
strip length of 50~$\mu$m and a closed steering-ring. 
The first measurement performed was a 1D scan repeated for different bias voltages.
The results are shown in Figure~\ref{fig13} for the strip and the steering-ring along with the layout to illustrate the 
the scan direction. The data are plotted in arbitrary units according to the normalisation with beam intensity 
variations.
The signal profile acquired from the strip (Figure~\ref{fig13a}) appears to be very well defined with a FWHM
of 17.2~$\mu$m, in close agreement with numerical simulations (Figure~\ref{fig5}). The signal profile has little 
dependence on bias voltage due to the thin active volume. The raw measured current is reported
for the maximum points of the measurements at 1V and 30V with an average value of roughly 4~$\mu$A.
The rather high current value is related to the large amount of photons hitting the device, 
$\sim$7.5$\times$10$^{10}$~ph./s for a synchrotron current of $\sim$80mA.
Confirming the findings of numerical simulations, a decrease in signal is observed in the middle of the strip,
where the highly doped n$^+$ implantation is present. 
The signal coming from the steering-ring (Figure~\ref{fig13b}) is complementary to the one of the strip. 
The depletion of the region outside of the ring is difficult for this bulk doping concentration. Signal generated 
in this undepleted region is therefore a lot lower than the one generated in the strip. This further confirms the 
successful shielding of each individual strip by the steering-ring.

\begin{figure}[!t]%
\centering 
\subfigure[]{\label{fig14a} 
	\includegraphics[width=.32\textwidth]{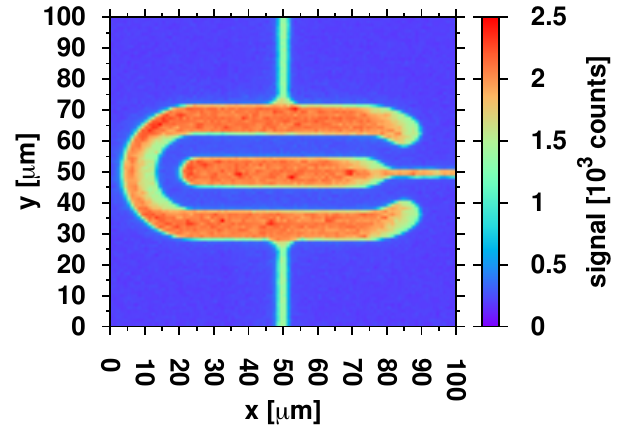}
}%
\subfigure[]{\label{fig14b} 
	\includegraphics[width=.32\textwidth]{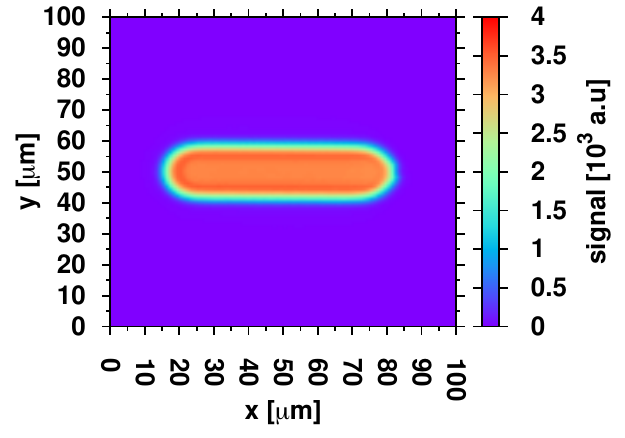}
}%
\subfigure[]{\label{fig14c}
	\includegraphics[width=.32\textwidth]{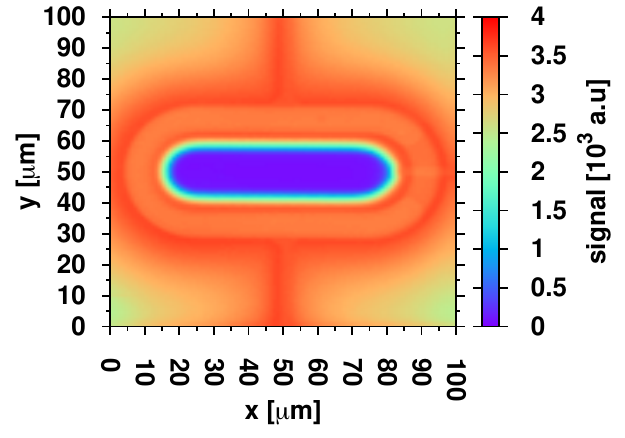}
}%
\caption{Bi-dimensional scan of a strip sensor from wafer A9 with a strip length of 50~$\mu$m and a closed 
steering-ring. The plots shown are the signal coming from the SDD measuring the aluminium fluorescence
(a) and the signals readout from the strip (b) and the steering-ring (c). The applied bias voltage was 30V.} 
\label{fig14}
\end{figure}

Bi-dimensional scans were also performed in the same operating conditions. 
The results of a measurement performed for a bias voltage of 30V are shown in Figure~\ref{fig14}. 
The total scanned area is 100$\times$100~$\mu$m$^2$. The signal of the fluorescence detector, indicating
were aluminium is present on the sensor surface is shown in Figure~\ref{fig14a}. 
The signal coming from the strip is very sharp and well defined (Figure~\ref{fig14b}). 
Using 2D scans provides a better understanding of the steering-ring operation 
(Figure~\ref{fig14c}). The inner region sees near zero signal, while under the ring itself and all around it, the signal 
is evident. A decrease in signal is observed in the corners of the image, caused by the lack of full depletion
in between two adjacent channels. High signal is also present under the vertical metal links, suggesting that 
some MOS field effects are present. No large differences in response were observed for devices with
different bulk doping concentrations.

\subsubsection{Signal evolution studies}
\begin{figure}[!t]%
\centering 
\subfigure[]{\label{fig15a} 
	\includegraphics[width=.45\textwidth]{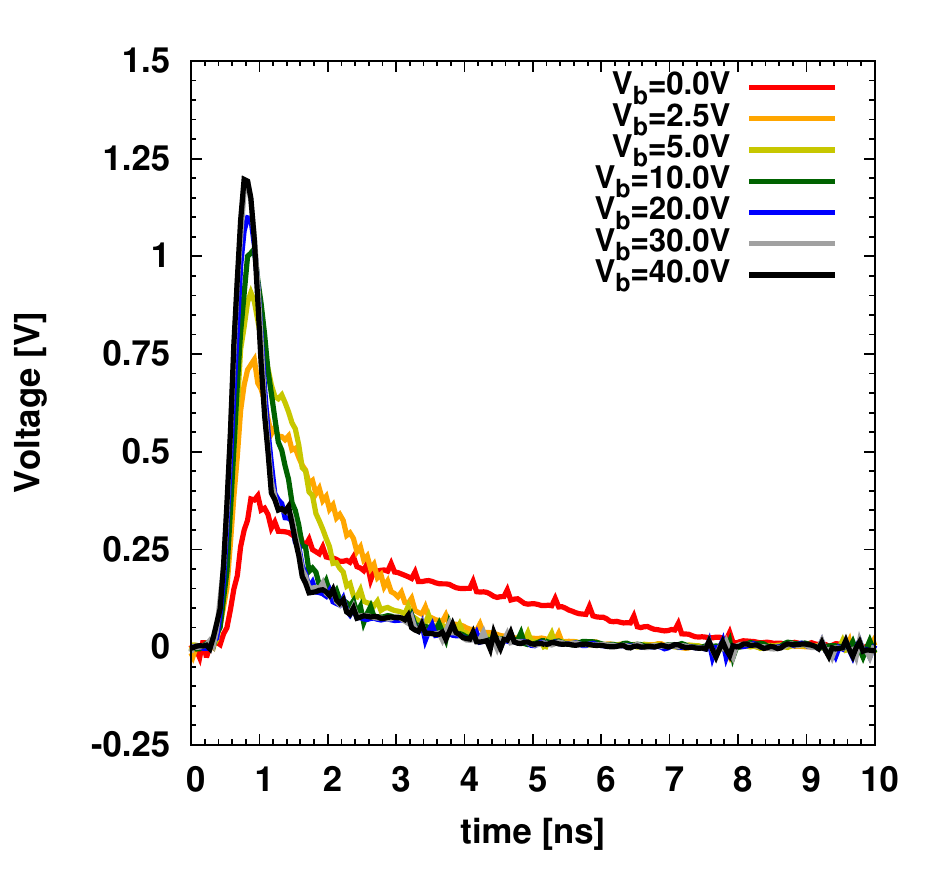}
}%
\subfigure[]{\label{fig15b} 
	\includegraphics[width=.45\textwidth]{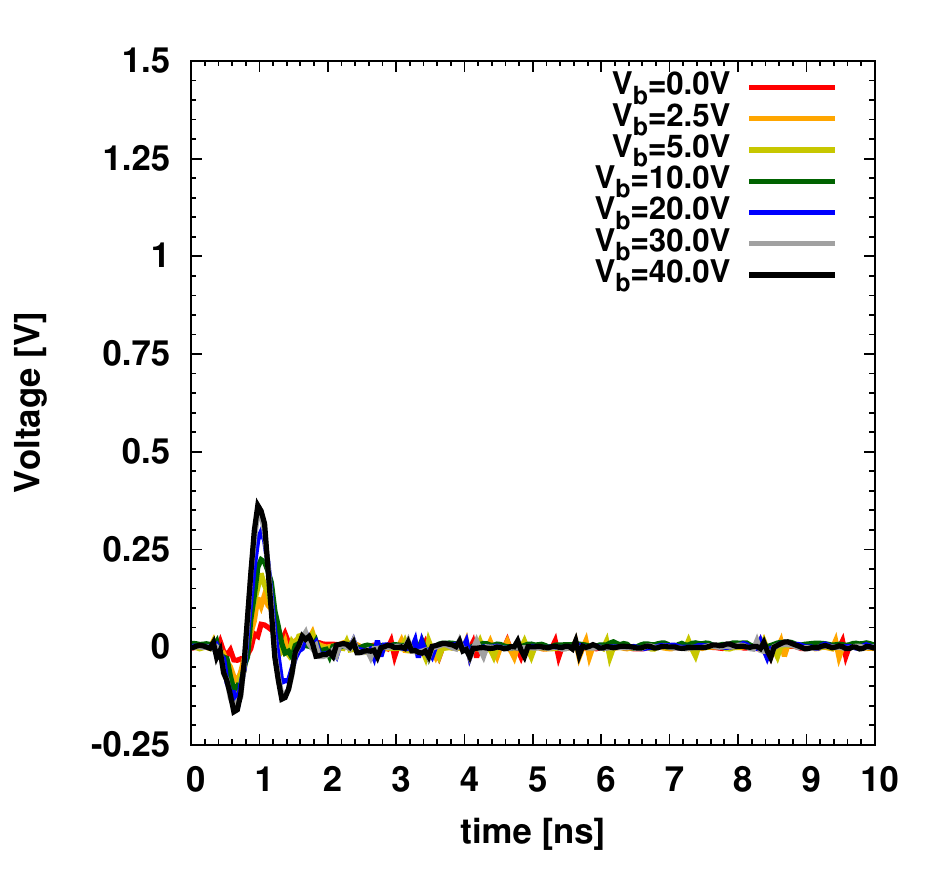}
}%
\caption{Acquired signals as a function of time and bias voltage for a strip (a) and for the steering-ring (b) with
the X-ray beam aligned on the strip.} 
\label{fig15}
\end{figure}

Another interesting measurement available at ID21 is the study of time evolution of the sensor signal.
When the synchrotron is operating in 16-bunch mode, a synchronisation signal is sent out to the beam line at 
every bunch. Using this timing signal to trigger the DSO it is possible to acquire the sensor current as a function
of time in different positions and for different bias voltages. As an example, a result of this type of measurement 
is reported in Figure~\ref{fig15}. With the beam aligned in the middle of one strip, the signal is acquired
and plotted as a function of time and bias voltage. In Figure~\ref{fig15a} it is possible to observe how low
bias voltages cause the sensor to be under depleted, resulting in a long signal "tail" reaching almost 10~ns
in duration, a rather long time considering that the device is only 10~$\mu$m thick. 
As the bias is increased, the amplitude of the signal increases, while the width is strongly reduced 
reaching roughly 1~ns at 40V.
The result is as expected although one would expect a even faster collection time for such a thin device.
Considering the bandwidth of the pre-amplifier (2.3~GHz) and the DSO (3~GHz),
the main limitation in signal evolution is related to the time constant $\tau = R_{in}C_{det}$, 
where $R_{in}$ is the DSO input resistance, C$_{det}$ is the detectors capacitance including also
parasitic capacitances due to sensor assembly and cabling. 
Considering $R_{in}=50~\Omega$ and $C_{det}\simeq30$~pF, the time constant 
would be 1.5~ns confirming its limiting role in signal evolution.  
On the contrary, the rising front is extremely fast, in the order of a few hundred picoseconds. 
The signal readout from the steering-ring is shown in Figure~\ref{fig15b}, 
where transient signals are observed but with no net integration value. This is 
attributed to induced signal, in agreement with Ramo's theorem \cite{Ramo,Eremin}.

\begin{figure}[!t]%
\centering 
\subfigure[]{\label{fig16a} 
	\includegraphics[width=.45\textwidth]{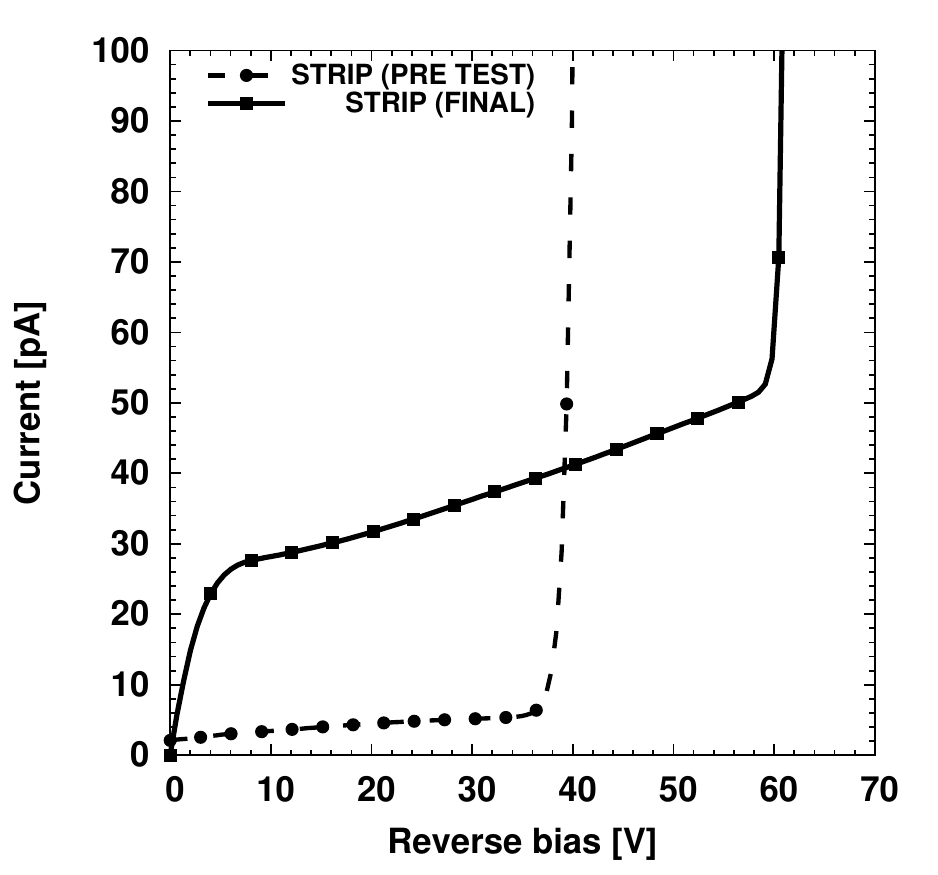}
}%
\subfigure[]{\label{fig16b} 
	\includegraphics[width=.45\textwidth]{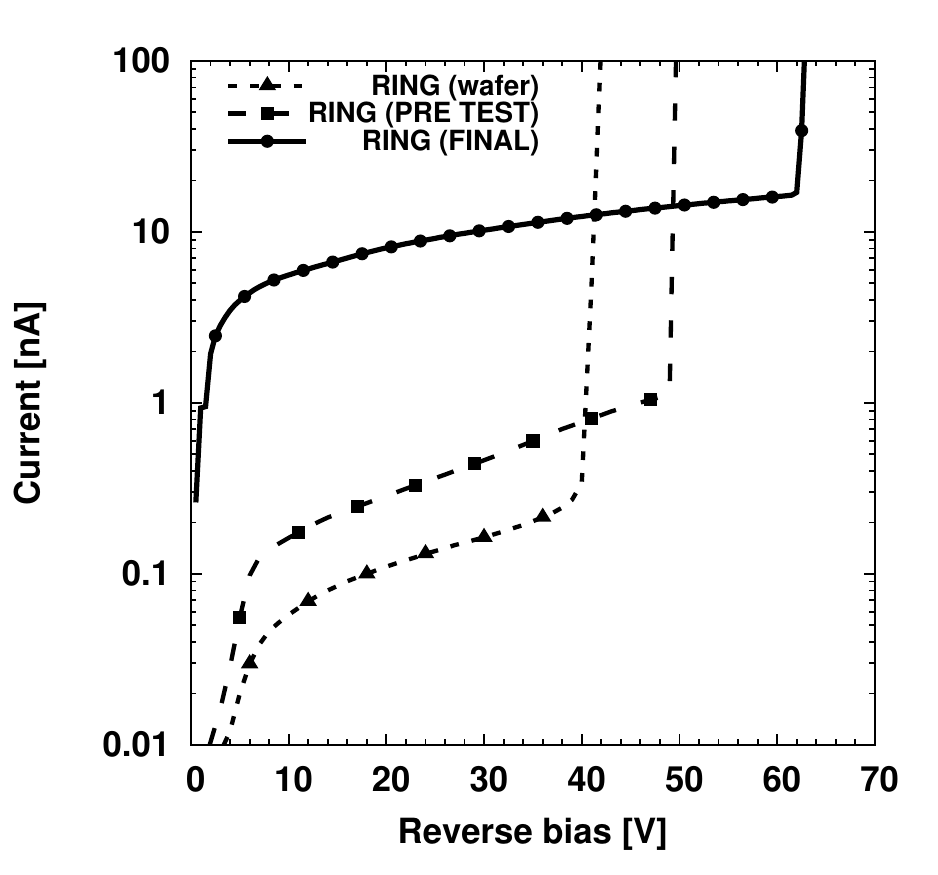}
}%
\caption{Current-Voltage measurements for the steering-ring (a) and for the strip (b) before and
after testing at ID21.} 
\label{fig16}
\end{figure}

As a final test, the reverse currents are re-measured after the testing to understand if any radiation damage 
is observed. The results are reported in Figure~\ref{fig16a} for the strip and in Figure~\ref{fig16b} for the
steering-ring. In both cases an increase in reverse current and breakdown voltage is observed on the final
measurement (solid line). In particular the steering-ring sees a total increase in current of almost one order
of magnitude and both contacts see an increase in breakdown of about 20~V. Both results are in agreement
with the expected consequences of radiation damage occurring at the interface between silicon and silicon oxide.
The increase in current is related to the increased surface recombination velocity and trapping, 
while the increase in breakdown voltage is a direct consequence of the charging of the surface oxide 
layers (positive charge) which typically results in an increased concentration of negative charge at the interface, 
slightly reducing the effective doping concentration of the p-spray implantation and thus increasing the 
maximum operating voltage of the sensor. 
No bulk damage is expected for X-rays of 7.2~keV. No reduction in signal was observed, indicating 
that the devices are relatively robust to this type of radiation. These results are not completely conclusive because
of the rather non-uniform irradiation field (the beam was focused on a small area for long periods of time
and not all the strips were irradiated in the same way). A more detailed irradiation study is planned
at ID17 to try to estimate the maximum dose these devices can withstand before failing.

\section{Improvements to design layout}\label{sec:layout-improvements}
After the described tests some modifications can be suggested
for the next generation of MRT beam monitoring devices. As discussed in 
subsection~\ref{subsec:Beam-tests at ESRF ID17}, having a fixed strip pitch of 100~$\mu$m is not the 
optimal solution due to the beam divergence at the location where the sensor is placed, in particular if the 
distance between the MSC and the sensor might not be possible to decrease.
In addition, the sensor with variable pitch geometry was demonstrated to provide much better alignment as well 
as beam profile measurements. Based on these experimental findings and studies through numerical simulations, 
the next generation of sensors will be in both standard and variable pitch configurations where the standard pitch 
would be 100.5~$\mu$m instead of 100~$\mu$m. This would result in a pitch of 402~$\mu$m in every 5 
strips which can compensate the beam divergence given a certain distance of the sensor from the MSC.
In addition, a new fabrication technology will be implemented, the so-called 3D technology \cite{Parker97},
originally proposed to be used in particle tracking detectors. 
Using 3D technology, the steering-ring can be fabricated as a deep doped trench that can ensure full isolation 
of the strip from the rest of the sensor.  
Furthermore, the experimental results also suggested that a standard strip length of 50~$\mu$m is so far 
the most suitable strip length for most operating conditions.
The new sensors are currently being fabricated and will be completed in late 2015.

\section{Conclusions}
A novel, thin silicon strip detector for beam monitoring as part of a Q\&A was designed, simulated 
and tested. The fabrication process was successful, with a high yield of functional devices. The electrical
characteristics of the sensors were found to be in good agreement with the numerical simulations.
The functional testing at ID17 in operation mode of the MRT demonstrated the capability of these new 
devices to measure the beam profile reliably provided that correct positioning and alignment is achieved. 
Additional insight into the device behaviour were obtained from experimental characterisation using an X-ray 
scanning microscope at ID21.  Remarkable results in terms of charge collection uniformity and interesting
proof of charge collection dynamics as a function of time and bias voltage were observed.
The experimental results suggested the sensors have good radiation hardness but further studies are required for 
an accurate estimation of the maximum operating dose for the sensors in particular under the operating conditions 
of the MRT.

\acknowledgments
This work was carried out within the 3DMiMic project, funded by the Research 
Council of Norway within the NANO2021 program.

We acknowledge the European Synchrotron Radiation Facility for provision of synchrotron radiation 
facilities and we would like to thank Herwig Requardt, Murielle Salom\'e and all the staff for assistance in 
using beamline ID17 and ID21.

Authors Lerch, Petasecca, Rosenfeld and Fournier acknowledge the support of the National Health and 
Medical Research Council (APP1017394) and the Australian Synchrotron (ISAP8728 and ISAP9209).

\end{document}